\documentclass[utf8,bibauthoryear]{aa}  
\usepackage{graphicx}
\usepackage{txfonts}
\usepackage{siunitx}
\usepackage[utf8]{inputenc}
\usepackage{natbib}
\usepackage{caption}
\usepackage{subcaption}
\usepackage{amsmath}
\usepackage{silence}
\usepackage{xcolor}
\usepackage{array}
\usepackage{multirow}
\usepackage{bm}
\usepackage[pdftex, hidelinks,draft]{hyperref}
\addto\extrasenglish{  
    
}
\defcitealias{2019A&A...632A.114S}{FS19}
\defcitealias{2017ApJ...841...30T}{PT17}
\defcitealias{10.1111/j.1365-2966.2011.18315.x}{H11}
\bibpunct{(}{)}{;}{a}{}{,}
\addto\extrasenglish{
}

\begin{document}

   \title{Exploring deep and hot adiabats as a potential solution to the radius inflation problem in brown dwarfs}
   \titlerunning{A Solution to the Radius Inflation Problem in Brown Dwarfs}
   \subtitle{Long-timescale models of the deep atmospheres of KELT-1b, Kepler-13Ab, and SDSS1411B }

   \author{F. Sainsbury-Martinez\inst{1}\thanks{\email{felix.sainsbury@cea.fr}} \and S. L. Casewell\inst{2} \and J. D. Lothringer\inst{3} \and M. W. Phillips\inst{4} \and P. Tremblin\inst{1} \fnmsep}

   \institute{Maison de la Simulation, CEA-CNRS-UPS-UVSQ, CEA Paris-Saclay\ \and School of Physics and Astronomy, University of Leicester, University Road, Leicester LE1 7RH, UK\ \and Department of Physics and Astronomy, Johns Hopkins University, Baltimore, MD 21210 USA\ \and School of Physics and Astronomy, University of Exeter, Exeter, EX4 4QL, UK}

   \date{Received 25 06 2021; accepted 17 09 2021}

  \abstract
   {} 
   {The anomalously large radii of highly irradiated gaseous exoplanets has remained a mystery for some time. One mechanism that has been suggested as a solution for hot Jupiters is the heating of the deep atmosphere via the vertical advection of potential temperature, resulting in increased internal entropy. In this work, we intend to explore whether this mechanism can also explain the observed brown dwarf radii trend: a general increase in the observed radius with irradiation, with an exception, however, for highly irradiated brown dwarfs orbiting white dwarfs.} 
  {We used a 3D global circulation model (GCM) known as DYNAMICO to run a series of long-timescale models of the deep atmospheres of Kepler-13Ab, KELT-1b, and SDSS1411B. These models allowed us to explore not only whether a stable advective adiabat can develop in this context, but also to consider the associated dynamics. }
   {We find that our brown dwarf models fall into two distinct regimes. First, Kepler-13Ab and KELT-1b both show signs of significant deep heating and, hence, are able to maintain adiabats that are hotter than 1D models predict. { On the other hand, SDSS1411B exhibits a much weaker downward heating profile that not only struggles to heat the interior under ideal conditions, but is highly sensitive to the presence of deep radiative dynamics.} }
   {We conclude that the vertical advection of potential temperature by large-scale atmospheric circulations constitutes a robust mechanism to explain the trend of increasing inflation with irradiation{. This includes an exception for highly irradiated brown dwarfs orbiting white dwarfs, which can be understood as occurring due to the role that increasing rotational influence plays in the context of mid-to-high latitude advective dynamics. Furthermore, when paired with a suitable parametrisation of the outer atmosphere irradiation profile, this mechanism alone could potentially provide a complete explanation for the observed levels of radius inflation in our brown dwarf sample. Finally, in order to confirm the validity of this explanation,} we suggest that this work should be followed by future studies of brown dwarfs atmospheres using next-generation, fully radiative GCMs. }


   \maketitle
%

\section{Introduction} \label{sec:introduction}
The anomalously large radii of highly irradiated gaseous exoplanets remains one of the key unresolved areas in our understanding of extrasolar giant planetary atmospheres. In recent years, however, this state of affairs has started to change, at least for highly irradiated Jupiter-like planets (i.e. hot Jupiters). In particular, a large number of different mechanisms have been suggested as possible explanations for the observed inflation (see \citealt{2014prpl.conf..763B} and \citealt{2021JGRE..12606629F}, for recent reviews of this topic), as well as the correlation between said inflation and the planetary irradiation \citep[for examples, see][]{Demory_2011,2011ApJ...729L...7L,2016ApJ...818....4L,2018A&A...616A..76S,2018AJ....155..214T}. However, much less work has gone into trying to understand the inflated radii of { lower-mass} brown dwarfs. At first glance, it seems reasonable to suggest that the same mechanism responsible for radius inflation in hot Jupiters should also operate within these brown dwarf atmospheres given how qualitatively similar they are. However, this is not guaranteed as not only must the inflation mechanism operate in an atmosphere in which the influence of gravity is significantly enhanced, it must also be able to explain the observed correlation between inflation and irradiation for brown dwarfs \citep{2020MNRAS.499.5318C}. This correlation is not simply a linear increase in radius with irradiation, but, rather, a combination of two trends: an increase in the observed radius with irradiation for brown dwarfs orbiting main sequence stars and a lack of inflation for very highly irradiated brown dwarfs orbiting hot white dwarfs (see \autoref{sec:obs}). We note that a third trend, namely, that of decreasing inflation with increasing giant planetary mass, has also been observed; however, \citet{2021ApJ...909L..16T} suggests that this effect ranks rather low in { importance} when compared to the link between irradiation and inflation.   \\

Recently, \citet[][hereafter \citetalias{2017ApJ...841...30T}]{2017ApJ...841...30T} and \citet[][hereafter \citetalias{2019A&A...632A.114S}]{2019A&A...632A.114S} explored in steady-state 2D{ or} long-timescale 3D models, respectively, a hot Jupiter inflation mechanism that naturally arises from first physical principles, namely: the formation of a hot, deep, and non-convective adiabat via the vertical transport of potential temperature. \\
Rather than developing via convective flows, the formation of this adiabatic region (i.e. region of constant potential temperature) in the deep atmosphere is instead driven by the vertical advection of potential temperature from the highly irradiated outer atmosphere to the deep atmosphere by large-scale dynamical motions, such as the equatorial super-rotating jet and its associated mass balancing flows. 
Here, it is almost completely homogenised by deep zonal and meridional flows and circulations. Critically this results in the `deep' temperature-pressure profile converging onto an adiabat at lower pressures than that at which the atmosphere might be expected to be unstable to convection. As a result, the outer atmosphere connects to a hotter internal adiabat than would be obtained from a standard 1D `radiative-convective' model potentially leading to a highly irradiated exoplanet having a larger radius than predicted. \citetalias{2017ApJ...841...30T} showed that this mechanism is able to reproduce the observed correlation between radius inflation and irradiation in the case of hot Jupiters.  \\

The purpose of this paper is to extend the work of \citetalias{2017ApJ...841...30T} and \citetalias{2019A&A...632A.114S} into the brown dwarf regime and explore whether the presence of a deep, hot, and non-convective adiabat { can also explain the radius inflation trend observed for these objects}. 
To that end, we performed a series of somewhat idealised, long-timescale, 3D global circulation model (GCM) calculations of three brown dwarfs: Kepler-13Ab, KELT-1b, and SDSS1411B, the latter of which shows no signs of radius inflation in observations and is in a very short orbit around a hot white dwarf. { These simulations will allow us to not only explore if a stable, inflating, advective adiabat is able to form, but also to investigate the flows, circulations, and fluxes associated with said development, or a lack thereof.}

The structure of this work is as follows: We start, in \autoref{sec:obs}, by detailing some key observations of brown dwarfs{. This} includes an overview of the mass-radius-irradiation relation for known transiting hot brown dwarfs as well as key observational parameters for the three brown dwarfs (and their host-stars) under consideration here: Kepler-13Ab, KELT-1b, and SDSS1411B. Next, in \autoref{sec:method}, we introduce the model used here to explore this inflation mechanism{. This} includes an overview of both the numerical scheme used in our 3D GCM (DYNAMICO), as well as the 1D models that we use to define and parametrise said models radiative dynamics{. Then,} in \autoref{sec:results}, we detail our results. We start by exploring the differences between models of each of our three brown dwarfs, after which we introduce a series of test cases designed to explore at what pressure (referred to as the horizontal convergence pressure) each brown dwarf model atmosphere might be expected to develop a deep, hot, advective adiabat (\autoref{sec:1000bar}). Having calculated a horizontal convergence pressure range for each brown dwarf, we then explore how robust each brown dwarfs advective adiabat is to the presence of deep radiative forcing on a similar (or slightly weaker) timescale to that seen in 1D `radiative-equilibrium' models (\autoref{sec:radiative_stability}). Finally, in \autoref{sec:flows}, we explore the flows, circulations, and fluxes which lead to the development, or lack thereof, of a radiatively stable deep adiabat for each of our brown dwarfs. We conclude, in \autoref{sec:conclusion}, by providing concluding remarks, discussing the implications of our results, and introducing suggestions for future computational studies of lower mass brown dwarf atmospheres using next-generation radiative 3D GCMs. 

\section{Observations of brown dwarfs: Kepler-13Ab, KELT-1b, and SDSS1411B} \label{sec:obs}
\begin{figure*}[btp] %
\begin{centering}
\includegraphics[width=0.95\textwidth]{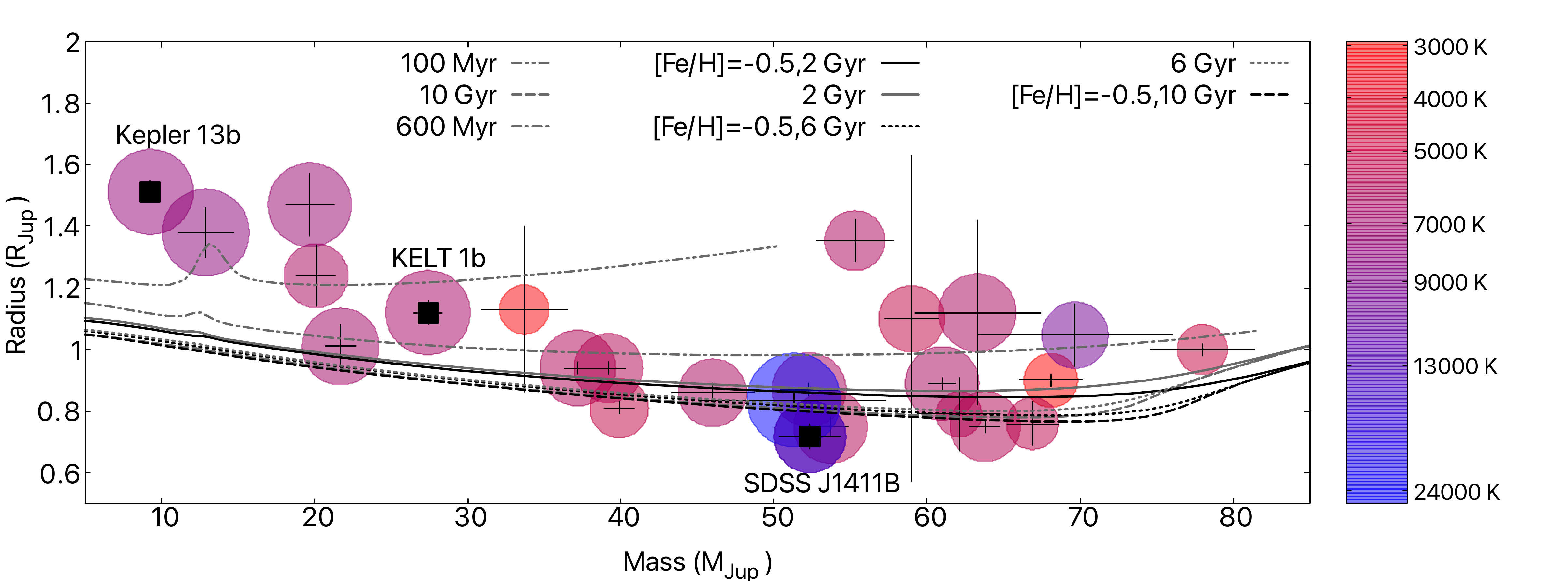}
\caption{ Mass-radius relation for 25 of the known highly irradiated brown dwarfs transiting either a main sequence star (\citealt{2020AJ....160...53C} and references therein) or a white dwarf \citep{2020MNRAS.497.3571C,2014MNRAS.445.2106L,2017MNRAS.471..976P}, with the brown dwarfs under consideration here clearly labelled. Additionally, the effective temperatures of the brown dwarf hosting stars are indicated by the colour of each brown dwarfs circle, while the size of each coloured circle is proportional to the total incident surface irradiation. Also shown are the Sonora-Bobcat evolutionary models of \citet{marley_mark_2018_1309035} for 100 Myr and 600 Myr at only solar metalicity, and at 2 Gyr, 6 Gyr, and 10 Gyr for both solar metallicities (grey) and low metallicities ([Fe/H]=-0.5 - black). \label{fig:MR_colour_2}  }
\end{centering}
\end{figure*}

To date, observations have revealed 25 highly irradiated brown dwarfs that transit a main sequence star (\citealt{2021AJ....161...97C} and references therein) and three that transit white dwarfs \citep{2020MNRAS.497.3571C,2014MNRAS.445.2106L,2017MNRAS.471..976P}. \autoref{fig:MR_colour_2}, in which we plot the mass-radius-irradiation relation for 25 of these transiting brown dwarfs, reveals a somewhat similar trend to that seen for highly irradiated Jupiter-like planets: a general increase in the observed radius with increasing irradiation, albeit with a significant reduction in inflation strength { towards higher} masses \citep{2021ApJ...909L..16T}. However, this trend is not universal with two of the most highly irradiated brown dwarfs included in \autoref{fig:MR_colour_2} showing little to no sign of radius inflation{; that is to say that the observations closely match the radii predicted by evolutionary models}. So, we consider what separates these brown dwarfs from their compatriots and we find that{ one} major factor appears to be that they are{ situated in short orbits around} white dwarfs.

To date, no mechanism has been suggested that can explain either the general trend of increasing inflation with irradiation (albeit tempered by mass{) or} the exception to this trend for very highly irradiated brown dwarfs orbing hot white dwarfs.\\ 

Here, we intend to investigate whether potential temperature advection{ can (via a simple physical mechanism) finally provide this explanation.  To that end, we intend to explore the steady state deep atmospheres of three representative brown dwarfs: Kepler-13Ab, KELT-1b, and SDSS1411B. Studies of Kepler-13Ab and KELT-1b reveals a significant discrepancy between the observed and theoretical radii for both brown dwarfs, albeit to a slightly smaller degree than might be expected for a similarly located hot Jupiter thanks to the attenuating effect of their comparatively higher masses \citep{2021ApJ...909L..16T}.
Whereas observations of SDSS1411b show little to no signs of radius inflation despite its very short orbit around a very hot white dwarf, resulting in a strong surface irradiation that would otherwise imply the presence radius inflation even when accounting for the attenuating effects of mass.} 

All 3 objects studied are assumed to be tidally locked due to their proximity to their host star. Host star and brown dwarf parameters can be found in \autoref{tab:star_params}.
\begin{table*}[htbp!]
\centering
\begin{tabular}{c|ccc}
 Parameter (units) & Kepler-13A &  KELT-1A  & SDSS-J1411 \\ \hline \hline
 Surface Temperature (\si{\kelvin})  & $7650\pm250$ &   $6518\pm50$  & $13000\pm300$ \\
 $v\sin i$ (\si{\kilo\meter\per\second})  & $78\pm15$ &   $56\pm2$  & N/A \\
 Mass ($M_{\sun}$) & $1.72\pm0.10$ & $1.324\pm0.026$ & $0.53\pm0.03$ \\
 Radius ($R_{\sun}$) & $1.71\pm0.04$ & $1.462^{+0.037}_{-0.024}$ & $0.0142\pm0.0006$ \\
 Metalicity ($[Fe/H]$) & 0.2 & $0.008\pm0.073$ & N/A \\
 Surface Gravity ($\log g$) & $4.2\pm0.5$ &$4.229^{+0.012}_{-0.019}$& $7.86\pm0.07$\\
 Age (Gyr) & $0.5\pm0.1$ & $1.75\pm0.25$ & >3\\
 \hline
 &Kepler-13b &  KELT-1b  & SDSS-J1411B \\ \hline \hline
 Mass$_{\rm BD}$ (M$_{\rm J}$)&9.28$\pm$0.16& 27.38$\pm$0.93&52.37$\pm$2.09\\
 Radius$_{\rm BD}$ (R$_{\rm J}$) &1.512 $\pm$0.035&$1.116^{+0.038}_{-0.029}$&0.716$\pm$0.039\\
 Surface Gravity ($\log g$) & 4.02 $\pm$0.03&$4.736^{+0.017}_{-0.025}$&5.342$\pm$0.1\\
 Semi-major axis (AU)&0.03641$\pm$0.00087 &0.02472$\pm$0.00039 & 0.00316$\pm$0.000046\\
 T$_{\rm dayside}$ (\si{\kelvin}) & $2750\rightarrow3490$ & $3340\pm110$ & $1730\pm70$ \\
 T$_{\rm nightside}$ (\si{\kelvin}) & $\sim2500$ & $1820^{+640}_{-1150}$ & $1540^{+90}_{-70}$ \\
 { T$_{\rm eq}$ @ Zero Albedo (\si{\kelvin})} & {$2570$} & {$2426$} & {$1367$}
\end{tabular}
  \caption{Parameters for both primary star and brown dwarf of Kepler-13Ab, KELT-1b, and SDSS1411B}
  \label{tab:star_params}
\end{table*}

\subsection{Kepler-13Ab}
Kepler-13Ab is a low-mass brown dwarf detected by the $Kepler$ space telescope \citep{2011ApJ...736...19B,2011ApJ...736L...4S}. It has an inflated radius of $1.512\pm0.035 \mathrm{R_{J}}$,  a mass of $9.28\pm0.16 \mathrm{M_{J}}$, and an orbital period of $\sim$1.76 days \citep{Esteves_2015}. The observations and models suggest that Kepler-13Ab has an average night-side (brightness) temperature of $\sim2500\textrm{K}$ \citep{Shporer_2014,Esteves_2015} and a day-side (brightness) temperature of between $2750\textrm{K}$ \citep{Shporer_2014} and $3490\textrm{K}$ \citep{Esteves_2015}. For our models, we take an average of these two results and set the day-night temperature difference to $600\textrm{K}$ at $0.01 \mathrm{bar}$. { Finally, calculations suggest that the zero albedo equilibrium temperature of this brown dwarf is $2570\textrm{K}$}. 

The spectra of Kepler-13Ab reveal an absorption feature due to water  \citep{2017AJ....154..158B}, whereas 1D models would suggest that any water feature should either be in emission (because of a temperature inversion) or muted (from an isothermal structure: see \citealt{2018ApJ...866...27L}). A number of explanations for this feature have been put forth (e.g. \citealt{2020ApJ...905..163L}), including the suggestion that the water absorption feature is seen because observations are probing the internal adiabat{. However, in 1D models, this requires that the internal temperature be increased from 1000~K, corresponding to a 1D radius of $\sim1.195\mathrm{R_{J}}$, to at least 1750K - an increase which is incompatible with the age and mass of Kepler-13Ab. Here, we investigate whether, instead, the vertical advection of potential temperature can explain such a hot interior (\autoref{sec:free}).}  \\

The host star, Kepler-13A, is a chromospherically moderately rapidly rotating ($v\sin i = 78\pm15\si{\kilo\meter\per\second}$) A0V star in a close ($a=0.410\mathrm{AU}$) triple system (consisting of Kepler-13A and the binary subsystem Kepler-13B and Kepler-13C), with an effective temperature of $7650\pm250\mathrm{K}$ (Table \ref{tab:star_params}: \citealt{Shporer_2014}). \\

\subsection{KELT-1b}
KELT-1b was is the first object detected by the Kilodegree Extremely Little Telescope-North (KELT-North) transit survey \citep{2012ApJ...761..123S}. It has an inflated radius of $1.116^{+0.038}_{-0.029} \mathrm{R_{J}}$ and a mass of $27.38\pm0.93 \mathrm{M_{J}}$, which is three times that of Kepler-13Ab. Its orbital period is $1.21$ days \citep{2012ApJ...761..123S}, slightly shorter than that of Kepler-13Ab. The day-side of KELT-1b has a blackbody (brightness) temperature of $3340\pm110\mathrm{K}$, while the night-side has a more poorly constrained (brightness) temperature of $1820^{+640}_{-1150}\mathrm{K}$ \citep{2020AJ....160..211B}. Due to the uncertainty in these values, our models use a day-night temperature difference of $1200\textrm{K}$ at $0.01 \mathrm{bar}$ (with additional tests run with a $1600\textrm{K}$ temperature difference - tests which reveal similar dynamics with just a slight increase in jet speed and deep heating rate). { Finally, calculations suggest that the zero albedo equilibrium temperature of this brown dwarf is $2426\textrm{K}$}.

Comparisons with the 1D atmospheric and stellar evolution models of \citet{2020A&A...637A..38P} suggest that in order for the models to match the observed radius, the internal temperature of the brown dwarf must be increased from $\sim700\mathrm{K}$ (corresponding to a 1D radius of $\sim0.97\mathrm{R_{J}}$) to $\sim1600\mathrm{K}$, an increase which would require that the brown dwarf is either significantly younger or more massive than observations of the system suggest. Again, we aim to investigate whether, instead, potential temperature advection can explain such a hot, deep adiabat in \autoref{sec:free}.\\

The host star, KELT-1A, is an F5V, with a slightly slower surface rotation ($v\sin i = 56\pm2\si{\kilo\meter\per\second}$) than Kepler-13A (Table \ref{tab:star_params}:\citep{2012ApJ...761..123S,2014ApJ...783..112B}). Due to its spectral type, KELT-1A is expected to have low levels or no activity \citep{2012ApJ...761..123S}, although  \citet{2021A&A...648A..71V} have detected light curve modulation, which may be due to surface spots. \\

\subsection{SDSS1411B}
SDSS1411B (also know as SDSS-J1411+2009) is a highly irradiated brown dwarf in a very short orbit ($\sim$ 2~hrs) around a white dwarf \citep{2013A&A...558A..96B,2014MNRAS.445.2106L,2018MNRAS.481.5216C}. It has a radius of $0.072\pm0.004 \mathrm{R_{\sun}}$, consistent with model results, and a mass of $0.050\pm0.002 \mathrm{M_{\sun}}$, the highest mass brown dwarf in our study. H band observations suggest that SDSS1411B has a night-side brightness temperature of $1540^{+90}_{-70}\textrm{K}$ and a day-side brightness temperature of $1730\pm70\textrm{K}$ \citep{2018MNRAS.481.5216C}, giving a day-night temperature difference of $200\textrm{K}$ at $0.01 \mathrm{bar}$. { Finally, calculations suggest that the zero albedo equilibrium temperature of this brown dwarf is $1367\textrm{K}$}. 

At first glance, it might appear that the lack of radius inflation seen here is because SDSS1411B is significantly cooler than either KELT-1b or Kepler-13Ab, however, this is not the case. Firstly, SDSS1411B is not significantly cooler than numerous inflated hot Jupiters, for instance, HD209458b, and, as \citet{2019A&A...632A.114S} show, this object exhibits significant radius inflation. Secondly, \citet{2018AJ....155..214T} suggests that planets at $\sim1500-1750\textrm{K}$ are often highly inflated and that the inflation efficiency (or, rather, the fraction of incident flux that goes in to inflating the planet) appears to peak at these temperatures. { Thus, this is why we are so interested in finding out what separates SDSS1411B from other inflated gaseous planets as well as finding out if the advective heating mechanism can reproduce this difference}.  \\

The primary, SDSS-J1411A, is a hydrogen rich (DA) white dwarf with an effective temperature of $13000\pm300\mathrm{K}$, a slightly below-average mass of $0.53\pm0.03\mathrm{M_{\sun}}$ and an age greater than $3\mathrm~{Gyr}$ (Table \ref{tab:star_params}: \citealt{2014MNRAS.445.2106L}).

\begin{table*}[htbp!]
\centering
\begin{tabular}{c|c|ccc}
 Quantity (units) & Description & Kepler-13Ab &  KELT-1b  & SDSS1411B \\ \hline \hline
 dt (\si{\second}) & Time-step & 120 &   120  & 30 \\
 $N_z$ & Number of Pressure Levels & 44 &   44  & 44 \\
 $d$ & Number of Sub-divisions & 30 &   30  & 30 \\
 $N^\circ$ & Angular Resolution & 2.5 &   2.5  & 2.5 \\
 $P_{top}$ (\si{\bar}) & Pressure at Top & $1 \times 10^{-4}$ &   $1 \times 10^{-4}$ & $1 \times 10^{-4}$ \\
 $P_{bottom}$ (\si{\bar}) & Pressure at Bottom & 1000 &   1000  & 1000 \\
 $g$ (\si{\meter\per\second\square}) & Gravity & 100 &   550  & 2000 \\
 $T_{int,1D}$ (\si{\kelvin}) & 1D Model Internal Temperature & 1000 &   700  & 800 \\
  $T_{d/n}$ (\si{\kelvin}) & Day-Night Temperature Contrast & 600 &  1200  & 200 \\
 $\Omega$ (\si{\per\second}) & Angular Rotation Rate & $4.12 \times 10^{-5}$ &   $6.0 \times 10^{-5}$  & $8.6 \times 10^{-4}$ \\
 $c_p$ (\si{\joule\per\kilo\gram\per\kelvin}) & Specific Heat &  23500 & 16446 & 14917 \\
 $\mathcal{R}$ (\si{\joule\per\kilo\gram\per\kelvin}) & Ideal Gas Constant & 3779.0 &   3779.0  & 3779.0
\end{tabular}
  \caption{Parameters for core models of Kepler-13Ab, KELT-1b, and SDSS1411B}
  \label{tab:core_params}
\end{table*}
\section{Method} \label{sec:method}
In order to investigate whether the advection of the potential temperature can explain the observed radius-irradiation-host{ trend observed for} highly irradiated brown dwarfs, we must explore the steady state dynamics of the deep atmospheres of said objects. However, even if we start close to the equilibrium state (a deep adiabat based on the horizontal convergence pressure-temperature point - \citetalias{2019A&A...632A.114S}), to reach a steady-state at all pressures { is a slow process. Typically, thanks to the longer timescales of deep dynamics (see \citealt{Rauscher_2010,Mayne_2014,2019A&A...632A.114S} etc.), these models require an order of magnitude more simulation time ($\sim300$ Earth years) than would be used for equivalent models which focus only upon reaching equilibrium in the outer atmosphere} (e.g. ten Earth years to reach day-side radiative equilibrium at pressures of $< 10$ bar for the brown dwarf WD0137 - \citealt{2020MNRAS.496.4674L}). 

This, along with the limits on available computational resources, precludes us from using any of the current generation of fully radiative GCMs (such as the MetOffice GCM, Exo-FMS, etc...) for our models. As such, we turned to the highly computationally efficient next-generation GCM DYNAMICO (\autoref{sec:dynamico_NS}), which uses a Newtonian Cooling approach to model the radiative forcing (\autoref{sec:newtonian_cooling}) that we parametrise using day-side, night-side, and equilibrium 1D models calculated using the PHOENIX code (\autoref{sec:1D}). 

\subsection{DYNAMICO} \label{sec:dynamico_NS}
DYNAMICO is a highly computationally efficient GCM that solves the primitive equation of meteorology ({see \citealt{Vallis17} for a review and \citealt{2014JAtS...71.4621D} for a more detailed discussion of the approach taken in DYNAMICO}) on a spherical grid \citep{gmd-8-3131-2015}. It is being developed as a next-generation dynamical core for Earth and planetary climate studies at the Laboratoire de Météorologie Dynamique and is publicly available\footnote{DYNAMICO is available at http://forge.ipsl.jussieu.fr/dynamico/wiki}. Recently it has been used by \citetalias{2019A&A...632A.114S}\footnote{Using a hot Jupiter patch to DYNAMICO available at https://gitlab.erc-atmo.eu/erc-atmo/dynamico\_hj.} to model the deep atmosphere of the hot Jupiter HD209458b over a very long-timescale, and by \citet{2020Icar..33513377S} to model the atmosphere of Saturn at a very high resolution. \\

In brief, DYNAMICO takes an energy-conserving Hamiltonian approach to solving the primitive equations of meteorology. This has been shown to be suitable for modelling the atmospheres of both hot Jupiters and brown dwarfs (see \citealt{2020SSRv..216..139S} and references therein), although it may not be valid in other planetary atmospheres, such as small Neptunes and super Earths, \citep{2019ApJ...871...56M}. Rather than the traditional latitude-longitude horizontal grid (which presents numerical issues near the poles due to singularities in the coordinate system; see the review of \citealt{WILLIAMSON2007} for more details), DYNAMICO uses a staggered horizontal--icosahedral grid (see \citealt{gmd-7-909-2014} for a discussion of the relative numerical accuracy for this type of grid) for which the total number of horizontal cells, $N,$ is defined by the number of subdivisions, $d,$ of each edge of the main spherical icosahedral\footnote{Specifically, to generate the grid we start with a sphere that consists of 20 spherical triangles (sharing 12 vertex, i.e. grid, points) and then we subdivide each side of each triangle $d$ times using the new points to generate a new grid of spherical triangles with $N$ total vertices. These vertices then form the icosahedral grid.}:
\begin{equation}
  N=10 d^2 + 2.
\end{equation}
In all the models considered here, we set the number of subdivisions to 30, which results in a total horizontal resolution of 9002 cells. This corresponds to an angular resolution of approximately $2.5^\circ$.

As for the vertical grid, DYNAMICO uses a pressure coordinate system whose levels can be defined by the user at runtime. In our models, this means 44 pressure levels that are linearly spaced in $\log\left(P\right)$ space. Finally, the boundaries of our simulations are closed and stress-free with zero-energy transfer (i.e. the only means of energy injection and removal are the Newtonian cooling relaxation scheme - described in \autoref{sec:newtonian_cooling} -  and the horizontal numerical dissipation required to stabilise the system - see below). We note that unlike some other GCM models of gaseous giants (e.g. \citealt{2009JAtS...66..579S,2013ApJ...770...42L,2019ApJ...883....4S,2021MNRAS.502.2198T}), we do not include an additional frictional (i.e. Rayleigh) drag scheme at the bottom of our simulation domain, instead relying on the hyperviscosity (see below), the extended deep atmosphere { (with $P_{\textrm{max}}=1000\textrm{bar}$)}, and the impermeable bottom boundary to stabilise the system. \\

As a consequence of the finite difference scheme used in DYNAMICO, artificial numerical dissipation must be introduced in order to stabilise the system against the accumulation of grid-scale numerical noise. This numerical dissipation takes the form of a horizontal hyperdiffusion filter with a fixed hyperviscosity and a dissipation timescale at the grid scale, labelled $\tau_{dissip}$, which serves to adjust the strength of the filtering (the longer the dissipation time, the weaker the dissipation).
It is important to point out that the hyperviscosity is not a direct equivalent of the physical viscosity of the planetary atmosphere, but can be viewed as a form of increased artificial dissipation that both enhances the stability of the model and somewhat accounts for sub-grid-scale dynamics. This approach is known as the large eddy approximation and has long been standard practice in both the stellar (e.g. \citealt{2005LRSP....2....1M}) and planetary (e.g \citealt{doi:10.1098/rsta.2008.0268}) atmospheric modelling communities.

In a series of benchmark cases, \citet{10.1111/j.1365-2966.2011.18315.x} have shown that both spectral and finite-difference-based dynamical cores which implement horizontal hyperdiffusion filters can produce differences of the order of tens of percent in the temperature and velocity fields when varying the dissipation strength. A similar sensitivity to the dissipation timescale was found and explored by \citetalias{2019A&A...632A.114S} for models of HD209458b. The solution enacted by \citetalias{2019A&A...632A.114S} was to calibrate the dissipation timescale by minimising unwanted small-scale numerical noise as well as replicating published benchmark results. As a result of these calibration tests, they eventually settled on a dissipation timescale of $\tau_{dissip}=2500\,\mathrm{s}$ for their low resolution runs and $\tau_{dissip}=1250\,\mathrm{s}$ for their mid resolution runs (which match the horizontal resolution of our brown dwarf models here). Due to the similarity of of our model setup, as well as the fact that we are both modelling highly irradiated gaseous planets, it would seem natural to use the same dissipation timescale for our models (since we lack benchmark models and detailed observations to directly compare against). However, as we expect at least some of our brown dwarf models to contain an even stronger super-rotating jet than in \citetalias{2019A&A...632A.114S} models of HD209458b, we chose to retain the weaker dissipation $(\tau_{dissip}=2500\,\mathrm{s})$ used in their low resolution model. We note that a limited number of test cases with $\tau_{dissip}=1250\,\mathrm{s}$ reveal very similar dynamics, including smaller differences in the jet speed than found when varying the temporal averaging period. \\

Finally, since DYNAMICO (like many other GCMs) does not include a dynamic time-step, a benchmark model was run for each brown dwarf in order to calculate the maximum stable time-step (which { both} balances the need to correctly reproduce the physics, { while also} minimising the computational cost of the long-timescale models). For both Kepler-13Ab and KELT-1b, this resulted in a 120 second time-step being chosen, whereas the more turbulent flows of SDSS1411B (\autoref{sec:flows}) necessitated a shorter 30 second time-step, quadrupling this model's relative computational cost. 

\subsection{Radiative forcing via Newtonian cooling} \label{sec:newtonian_cooling}

\begin{table*}[htbp!]
\centering
\small
\begin{tabular}{cc|ccccc}
\multicolumn{2}{c}{Model} & \multicolumn{5}{|c}{$\tau_{rad}$ Profile Interpolation Points $\left(\log\left(\frac{\tau}{1 \textrm{sec}}\right),\frac{P}{1 \textrm{bar}}\right)$}                                                                \\ \hline
\multirow{10}{*}{Kepler-13Ab} & 0.6 bar - 1D Match & $(3.5,3\times10^{-3})$ & $(3.8,5\times10^{-2})$ &$(5.1,0.6)$& $(\infty,\infty)$ & N/A  \\
 & 1 bar core & $(3.5,3\times10^{-3})$ & $(3.8,5\times10^{-2})$&$(5.5,1)$& $(\infty,\infty)$ & N/A  \\
                  & 1 bar - A&        $(3.5,3\times10^{-3})$           &   $(3.8,5\times10^{-2})$                & $(5.5,1)$&$(6.2,8)$&$(13,1000)$  \\
                  & 1 bar - B&        $(3.5,3\times10^{-3})$          &   $(3.8,5\times10^{-2})$                &     $(5.5,1)$              &$(6.6,8)$&$(14.5,1000)$  \\ 
                  & 1 bar - C&         $(3.5,3\times10^{-3})$         &    $(3.8,5\times10^{-2})$               &      $(5.5,1)$             &$(8.5,30)$&$(18,1000)$  \\
                  & 4.3 bar core &  $(3.5,3\times10^{-3})$& $(3.8,5\times10^{-2})$& $(6.0,4.3)$ & $(\infty,\infty)$ &  N/A \\
                  & 4.3 bar - A&   $(3.5,3\times10^{-3})$                &          $(3.8,5\times10^{-2})$         &    $(6.0,4.3)$               &$(13.25,1000)$& N/A \\
                  & 4.3 bar - B&   $(3.5,3\times10^{-3})$               &       $(3.8,5\times10^{-2})$            &      $(6.0,4.3)$            &$(14.5,1000)$& N/A \\ 
                  & 4.3 bar - C&    $(3.5,3\times10^{-3})$              &       $(3.8,5\times10^{-2})$            &       $(6.0,4.3)$           &$(16.5,1000)$& N/A \\
                  & 1000 bar&      $(3.5,3\times10^{-3})$            &         $(3.8,5\times10^{-2})$          &        $(6.0,4.3)$          &$(6.5,10)$& $(13.5,1000)$  \\ \hline
\multirow{10}{*}{KELT-1b} &1 bar& $(3.3,5\times10^{-3})$ & $(3.5,0.2)$ &$(5.2,1)$ & $(\infty,\infty)$&  N/A \\
 &10 bar core/1D Match& $(3.3,5\times10^{-3})$ & $(3.5,0.2)$ &$(7.0,10)$ & $(\infty,\infty)$&  N/A \\
                  & 10 bar - A&   $(3.3,5\times10^{-3})$                &          $(3.5,0.2)$         &$(7.0,10)$& $(7.4,100)$&$(11,1000)$ \\
                  & 10 bar - B&     $(3.3,5\times10^{-3})$             &         $(3.5,0.2)$          &$(7.0,10)$&$(12,1000)$ & N/A\\ 
                  & 10 bar - C&       $(3.3,5\times10^{-3})$           &        $(3.5,0.2)$           &$(7.0,10)$&$(13,1000)$ & N/A \\ 
 &40 bar core& $(3.3,5\times10^{-3})$& $(3.5,0.2)$& $(7.4,40)$ & $(\infty,\infty)$&  N/A \\
                  & 40 bar - A&            $(3.3,5\times10^{-3})$       &      $(3.5,0.2)$             &$(6.9,8)$& $(7.5,100)$&$(10,1000)$ \\
                  & 40 bar - B&       $(3.3,5\times10^{-3})$           &      $(3.5,0.2)$             &$(6.9,8)$& $(9.5,100)$&$(12,1000)$  \\ 
                  & 40 bar - C&       $(3.3,5\times10^{-3})$           &     $(3.5,0.2)$              &$(6.9,8)$&$(9.5,100)$&$(14,1000)$  \\ 
                  & 1000 bar&         $(3.3,5\times10^{-3})$          &      $(3.5,0.2)$             &$(7.4,40)$&$(7.5,100)$& $(10.5,1000)$  \\\hline
\multirow{2}{*}{SDSS1411B}& 100 bar core&$(1.5,1\times10^{-3})$&$(3.75,1\times10^{-2})$&$(4.15,5)$ &$(7,100)$&$(\infty,\infty)$  \\
                 & 100 bar - A&           $(1.5,1\times10^{-3})$        &      $(3.75,1\times10^{-2})$             &        $(4.15,5)$           &      $(7,100)$             &  $(10,1000)$
\end{tabular}
  \caption{Newtonian cooling timescales for models of Kepler-13Ab, KELT-1b, and SDSS1411B}
  \label{tab:newtonian_cooling}
\end{table*}

\begin{table*}[htbp!]
\centering
\small
\begin{tabular}{cc|cccccc}
\multicolumn{2}{c}{Model} & \multicolumn{6}{|c}{$T_{eq}$ Profile Interpolation Points $\left(\frac{T_{eq}}{1 \textrm{K}},\frac{P}{1 \textrm{bar}}\right)$}                                                                \\ \hline
\multirow{4}{*}{Kepler-13Ab} & 0.6 bar - 1D Match & $(4150,5\times10^{-4})$ & $(2950,1\times10^{-2})$ & $(2500,0.1)$ & $(2650,0.6)$ &  N/A&   \\
 & 1 bar & $(4150,5\times10^{-4})$ & $(2950,1\times10^{-2})$ & $(2500,0.1)$ & $(2325,1)$ &  N/A&   \\
& 4.3 bar &$(4150,5\times10^{-4})$ &  $(2950,1\times10^{-2})$& $(2500,0.1)$ & $(2350,1)$& $(2470,4.3)$& N/A   \\
                  & 1000 bar&  $(4150,5\times10^{-4})$                 &      $(2950,1\times10^{-2})$             &      $(2500,0.1)$             &$(2350,1)$&$(3100,10)$& $(5400,1000)$  \\\hline
\multirow{5}{*}{KELT-1b} &1 bar& $(3930,5\times10^{-3})$ & $(2950,1\times10^{-2})$ &$(2350,1)$ &N/A&&   \\
&10 bar& $(3930,5\times10^{-3})$  & $(2950,1\times10^{-2})$ & $(2350,1)$&$(2220,10)$&N/A&   \\
&10 bar - 1D Match&$(3930,5\times10^{-3})$& $(2950,1\times10^{-2})$& $(2350,1)$& $(2450,10)$&N/A&   \\
&40 bar&$(3930,5\times10^{-3})$& $(2950,1\times10^{-2})$& $(2350,1)$& $(2260,10)$&$(2300,40)$& N/A   \\
                  & 1000 bar&      $(3930,5\times10^{-3})$             &     $(2950,1\times10^{-2})$              &   $(2350,1)$                &$(2260,10)$&$(3000,100)$& $(4400,1000)$ \\\hline
SDSS1411B& Core&$(4340,1\times10^{-3})$&$(1550,2\times10^{-2})$&$(1150,0.2)$ &$(1750,100)$&& 
\end{tabular}
  \caption{Equilibrium temperature profiles for models of Kepler-13Ab, KELT-1b, and SDSS1411B}
  \label{tab:equilbirum_temperature}
\end{table*}
\begin{figure*}[htbp!] %
\begin{centering}
\begin{subfigure}{0.45\textwidth}
\begin{centering}
\includegraphics[width=0.99\columnwidth]{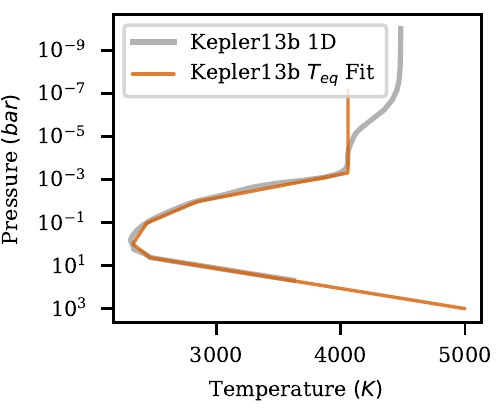}
\caption[]{Kepler-13Ab - Equilibrium 1D temperature profile fit  \label{fig:PT_K13} }
\end{centering}
\end{subfigure}
\begin{subfigure}{0.45\textwidth}
\begin{centering}
\includegraphics[width=0.99\columnwidth]{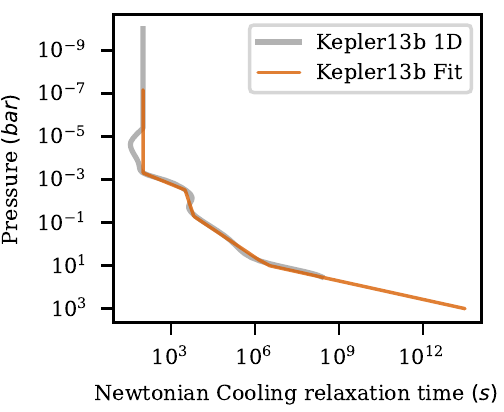}
\caption[]{Kepler-13Ab - Radiative forcing timescale fit \label{fig:timescale_K13} }
\end{centering}
\end{subfigure}
\begin{subfigure}{0.45\textwidth}
\begin{centering}
\includegraphics[width=0.99\columnwidth]{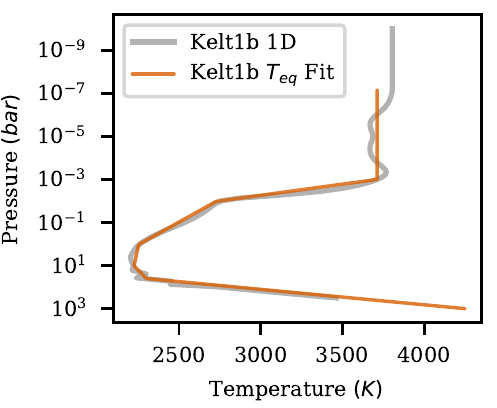}
\caption[]{KELT-1b - Equilibrium 1D temperature profile fit  \label{fig:PT_KELT} }
\end{centering}
\end{subfigure}
\begin{subfigure}{0.45\textwidth}
\begin{centering}
\includegraphics[width=0.99\columnwidth]{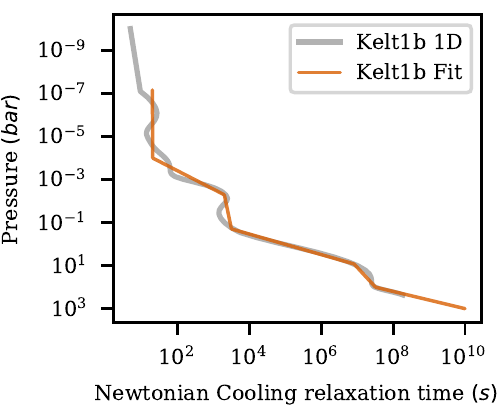}
\caption[]{KELT-1b - Radiative forcing timescale fit \label{fig:timescale_KELT} }
\end{centering}
\end{subfigure}
\begin{subfigure}{0.45\textwidth}
\begin{centering}
\includegraphics[width=0.99\columnwidth]{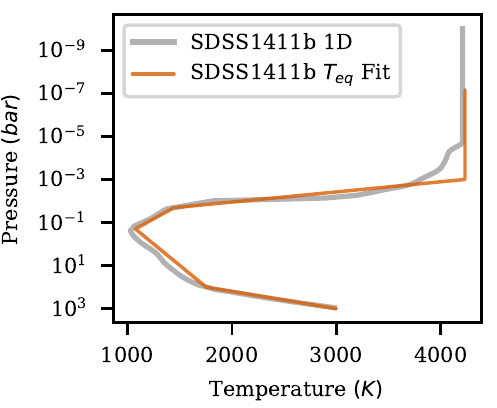}
\caption[]{SDSS1411B - Equilibrium 1D Temperature Profile Fit  \label{fig:PT_SDSS} }
\end{centering}
\end{subfigure}
\begin{subfigure}{0.45\textwidth}
\begin{centering}
\includegraphics[width=0.99\columnwidth]{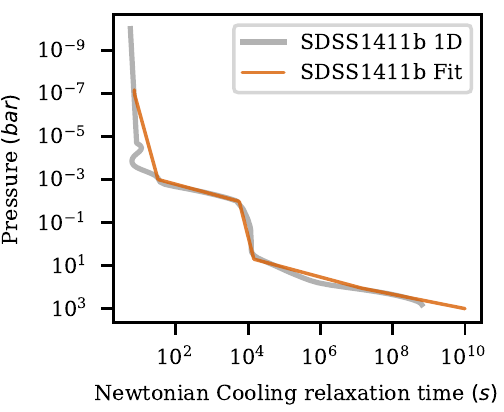}
\caption[]{SDSS1411B - Radiative forcing timescale fit \label{fig:timescale_SDSS} }
\end{centering}
\end{subfigure}
\caption{ Fits to the 1D equilibrium temperature-pressure ($T_{eq}$ - left) and Newtonian cooling relaxation timescale ($\tau_{rad}$ - right) profiles, which form the basis of the profiles used in our models of Kepler-13Ab (top), KELT-1b (middle), and SDSS1411B (bottom). In all cases, the equilibrium 1D profile is shown in light grey, and the linear in $\log\left(P\right)$ space fit is shown in orange. Note that depending upon the model parameters, including the pressure at which day-night temperature differences go to zero and presence of deep radiative dynamics, these fits may be truncated or modified as required. For full details of the exact fit used in each model, see both \autoref{tab:newtonian_cooling} and \autoref{tab:equilbirum_temperature}. } \label{fig:PT_and_timescale}
\end{centering}
\end{figure*}
In our simulations of brown dwarf atmospheres using DYNAMICO, we do not directly model either the incident thermal radiation on the day-side or the thermal emission on the night-side of the exoplanet. This would be prohibitively computationally expensive for the large array of long-timescale simulations we present here. Instead we use a simple thermal relaxation scheme to model these radiative effects, with a spatially varying equilibrium temperature profile, $T_{eq}$, and a radiative relaxation timescale, $\tau_{rad}$, that increases with pressure throughout the outer atmosphere. Specifically, we model the radiation by adding a source term to the temperature evolution equation that takes the form 
\begin{equation}
\frac{\partial T\left(P,\theta,\phi\right)}{\partial t} = - \frac{T\left(P,\theta,\phi\right)-T_{eq}\left(P,\theta,\phi\right)}{\tau_{rad}\left(P\right)} \,.
\end{equation}
This method, known as Newtonian cooling, has long been applied within the 3D GCM exoplanetary community (i.e. \citet{2002A&A...385..156G}, \citet{Showman_2008}, \citealt{Rauscher_2010}, \citet{2011ApJ...738...71S}, \citealt{2014GMD.....7.3059M}, \citealt{GUERLET2014110}, \citealt{Mayne_2014}, or \citetalias{2019A&A...632A.114S}), although it is gradually being replaced by coupling with simplified but much more computationally expensive radiative transfer schemes (e.g. \citealt{Showman_2009}, \citealt{2012ApJ...750...96R} or \citealt{2016A&A...595A..36A}) due to its many limitations { (particularly when studying the outer atmosphere)}, some of which we discuss in \autoref{sec:1000bar}.

For the majority of our models, $T_{eq}\left(P,\theta,\phi\right)$ is { calculated using the combination of a} simple linear in $\log\left(P\right)$ space fit to a 1D equilibrium model (see \autoref{sec:1D} for details about how these 1D models where calculated) { with a }pressure dependent day-night temperature difference:
\begin{align}
T_{eq}\left(P,\theta,\phi\right) &= T_{eq-1D}\left(P\right) - \frac{\Delta{T}(P)}{2} \notag \\
&+ \Delta{T}(P) \cos\left(\theta\right)
\max \left[ 0, \cos (\phi - \pi) \right] \, ,\\
\Delta T (P) &=\left\{ \begin{array}{ll}
  \Delta T_0 & \textrm{if } P<P_{low} \\
 \Delta T_0 \log (P/P_{low}) & \textrm{if } P_{low}  < P < P_{high} \\
 0 & \textrm{if } P > P_{high}  \end{array}
\right. \, , \label{eq:deltaT}
\end{align}
where $\Delta T_0$ is based on observations (see \autoref{sec:obs}) and remains constant for all models of a particular brown dwarf { (see \autoref{tab:core_params})}, $P_{low}=0.01$ bar for all models, and both $P_{high}$ and $T_{eq-1D}\left(P\right)$, are model dependent. { $P_{high}$ is simply the pressure at which the day-night temperature difference goes to zero (which we also define as the horizontal convergence pressure - i.e. it is the pressure at which we expect mixing to horizontally homogenise the atmosphere). As for $T_{eq-1D}\left(P\right),$ not only does this depend upon the horizontal convergence pressure (i.e. $P_{high}$), it also depends upon both the brown dwarf in question as well as the presence of any }deep radiative dynamics (which we explore via isothermal deep forcing - see \autoref{sec:radiative_stability}). The full $T_{eq-1D}\left(P\right)$ profiles that form the basis for each models individual profile are plotted on the left-hand side of \autoref{fig:PT_and_timescale}, { and} a full parametrisation of each models individual $T_{eq-1D}\left(P\right)$ profile can be found in \autoref{tab:equilbirum_temperature}.

Likewise, $\tau_{rad}$ is represented by a linear in $\log\left(P\right)$ fit to a 1D equilibrium model for each brown dwarf (see the right-hand side of \autoref{fig:PT_and_timescale}), although larger differences exist between the profile used in different models here, since adjusting $\tau_{rad}$ allows for a simple means to control the relative strength of the radiative forcing (in the deep atmosphere). In our `core' models, we used an infinite timescale to explore the formation of the adiabat deep profile by potential temperature advection alone, then we added a deep forcing (towards a deep isotherm) and verify how strong it needs to be to destroy the deep adiabatic structure that would be responsible for any observed radius inflation (in the spirit of what has been done in \citetalias{2019A&A...632A.114S}; see \autoref{sec:radiative_stability}).

\subsection{1D models of brown dwarf atmospheres} \label{sec:1D}

The 1D models which are parametrised as part of the Newtonian Cooling approach to radiative dynamics in our 3D models have been calculated using PHOENIX \citep{1997ApJ...483..390H,1999ApJ...512..377H,2001ApJ...556..885B,2011ApJ...733...65B}. {  PHOENIX is a }well-tested self-consistent atmospheric model that has been used to study the atmospheres of stellar and sub-stellar objects for decades, and which has previously been adapted to model the atmosphere of brown dwarfs \citep[e.g. ][]{2011ASPC..448...91A,2020ApJ...905..163L}.

In brief, PHOENIX works by iteratively calculating pressure-temperature profiles for a given 64-layer { P-T} structure on an log-spaced optical depth grid that extends from $\tau=10^{-8}$ to $\tau=10^{2.5}$ (which generally corresponds to pressures of between around $10^{-6}$ to $10^{2}$ bar). To start, the model first calculates chemical equilibrium using solar metalicity elemental abundances \citep{2006CoAst.147...76A} and by assuming that the atmosphere is in local thermodynamic equilibrium (LTE - i.e. collisional effects dominate the determination of the atomic and molecular compositions). Next, the opacity is calculated for each point using a wavelength grid that extends from 10 to $10^{6}$ Ångström with direct-Opacity-Sampling (dOS - \citealt{Schweitzer_2000}). The model includes over 130 different molecular opacity sources, plus both the atomic opacity for elements up to and including uranium, as well as many continuous opacity sources, including collision-induced-absorption and H- opacity. Radiative dynamics are then calculated including any irradiation by the host star which is defined using the closest matching stellar spectrum. We note that for the main sequence stars, we used the PHOENIX models from \citealt{2013A&A...553A...6H} and for the white dwarf, SDSS1411, we used the LTE DA white dwarf model from \citealt{2010MmSAI..81..921K}. Finally,  this can be used to calculate the vertical flux  and update the pressure-temperature structure, using a modified Unsöld-Lucy method \citep{2003ASPC..288..227H}, thus bringing the model slightly closer to radiative-equilibrium. This iterative process continues until the maximum temperature change for any layer is less than 0.5 K. \\

For each of the { brown dwarfs under consideration}, we calculated three 1D models: an equilibrium profile, a sub-stellar point profile, and a non-irradiated profile{. We note that the latter two of these profiles were used to represent a highly irradiated day-side and a non-irradiated night-side, respectively, in \autoref{sec:1000bar}, whilst the equilibrium profiles form the basis of the rest of our models.} \\
On top of the external irradiation by the host star (for the equilibrium and sub-stellar point profiles), each 1D model also requires that we set an internal temperature for the brown dwarf. { Following \citet{Baraffe_95}, we selected suitable warm internal temperatures that should be representative of the uninflated planets and checked for consistency against the models of \citet{2020A&A...637A..38P}. We note that we ended up using a slightly hotter internal temperature for Kepler-13Ab than the aforementioned models suggest, since the model was originally run as part of a test to see if 1D models alone could explain the atmosphere water feature of Kepler-13Ab. However, this should not be a significant problem because, in the way our models are setup, the exact value of the internal temperature has little effect on our primary results. This is because we primarily use the 1D profiles to set the forcing in the outer atmosphere, a region in which $T_{int}$ has little to no impact due to the local dominance of stellar radiative heating on the atmospheric dynamics. However, we still chose to use suitable values so as to both minimise errors and to ensure the accuracy of the day-night models to calculate the horizontal convergence pressures. \\}
For the brown dwarfs under consideration here, we set $T_{int} = 700K$ for KELT-1b, $T_{int} = 1000K$ for Kepler-13Ab, and $T_{int} = 800K$ for SDSS1411B. 
Finally, the two irradiated profiles for each brown dwarf include heat redistribution (with $f=0.25$ and $f=1.0$ for the equilibrium and sub-stellar point profiles respectively).  Using these internal temperatures, the planetary parameters (given in \autoref{tab:core_params}), and suitable stellar spectra, we are then able to calculate the aforementioned T-P and radiative timescale profiles.  We plot the resulting equilibrium profiles in \autoref{fig:PT_and_timescale} { and} the full set of 1D models can be found as part of the opendata archive\footnote{https://www.erc-atmo.eu/?page\_id=322}. 

\section{Results} \label{sec:results}
In order to explore the deep atmospheres of Kepler-13Ab, KELT-1b, and SDSS1411B, we have run a large array of models of each brown dwarf designed to explore not only at what pressure the deep adiabat might be able to develop (\autoref{sec:1000bar}), but also how stable this advective adiabat is against deep radiative dynamics (driven by Newtonian cooling - \autoref{sec:radiative_stability}), as well the link between its formation and the outer and deep atmosphere flows and circulations (\autoref{sec:flows}). The base parameters for each set of brown dwarf models are given in \autoref{tab:core_params}, and details of each models individual equilibrium temperature ($T_{eq}$) and radiative timescale ($\tau_{rad}$) profiles are given in \autoref{tab:newtonian_cooling} and \autoref{tab:equilbirum_temperature}, respectively. \\

\subsection{Calculating the radiative and advective boundaries for Kepler-13Ab and KELT-1b} \label{sec:1000bar}
\begin{figure}[htbp!] %
\begin{centering}
\begin{subfigure}{0.875\columnwidth}
\begin{centering}
\includegraphics[width=0.875\columnwidth]{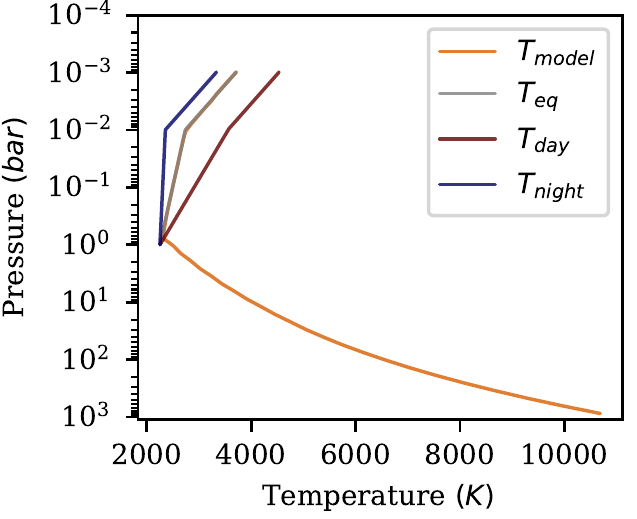}
\caption[]{KELT-1b - 1 bar convergence pressure \label{fig:Kelt1b_T_P_0_6} }
\end{centering}
\end{subfigure}
\begin{subfigure}{0.875\columnwidth}
\begin{centering}
\includegraphics[width=0.875\columnwidth]{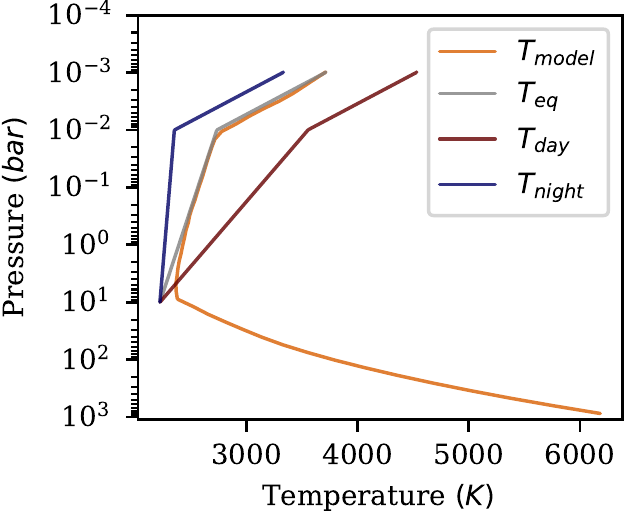}
\caption[]{KELT-1b -  10 bar convergence pressure \label{fig:Kelt1b_T_P_10}  }
\end{centering}
\end{subfigure}
\begin{subfigure}{0.875\columnwidth}
\begin{centering}
\includegraphics[width=0.875\columnwidth]{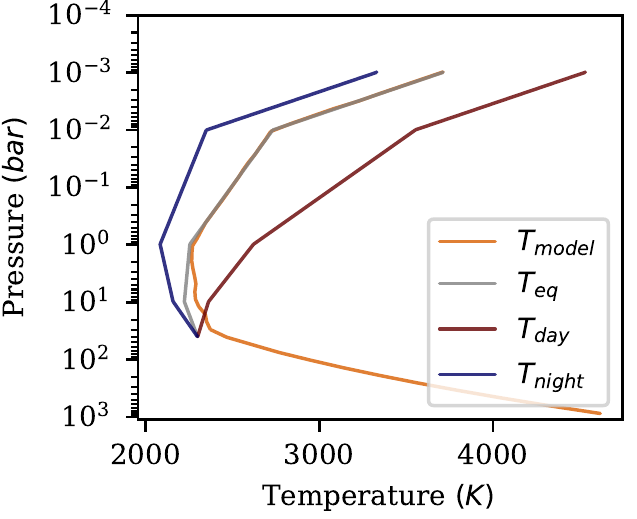}
\caption[]{KELT-1b - 40 bar convergence pressure \label{fig:Kelt1b_T_P_40}  }
\end{centering}
\end{subfigure}
\caption[]{ Equatorially averaged (i.e. the zonal-mean at the equator) T–P profiles (orange) for three models of KELT-1b with different horizontal convergence pressures ($P_{converge}$), and no radiative forcing, via Newtonian cooling, for $P>P_{converge}$. From top to bottom the convergence pressure of each model is 1 bar, 10 bar, and 40 bar, respectively{. In these plots, }the equilibrium Newtonian cooling profile is shown in grey, the day-side profile is shown in red, and the night-side profile in shown in blue.\label{fig:KELT_T_P_comp} } 
\end{centering}
\end{figure}
When modelling HD209458b, \citetalias{2019A&A...632A.114S} had a relatively easy time setting the 3D atmospheric temperature-pressure (T-P) profile of their models thanks to the wealth of data available for this well explored exoplanet (see, for example, \citealt{2005A&A...436..719I,10.1111/j.1365-2966.2011.18315.x,Mayne_2014}). However this is not the case for the brown dwarfs under consideration here. These objects have undergone much less study and so there exists no consensus on what form the 3D atmospheric T-P profiles should take. Fortunately observations can help to constrain the low pressure surface temperature profile, including the day-night temperature contrast at the top of our models (see \autoref{sec:obs}){. However, the same cannot be said for the deeper regions of the atmosphere, particularly because it is in these regions that we expect advective heat transport to play a critical role, leading to significant changes in the local dynamics when compared to the 1D models.}
Yet it is the structure of this `mid' atmosphere ($\sim0.1\rightarrow\sim10/100 \mathrm{bar}$) that is essential to improving our understanding of where a deep adiabat might be able to develop. \citetalias{2019A&A...632A.114S} showed that the formation of a deep adiabat is strongly linked to the horizontal homogenisation of the deep atmosphere.

In \autoref{sec:free}, we show that by simply adjusting the horizontal convergence pressure (and setting the deep radiative effects to be dynamically unimportant), we are able to form a deep adiabat at almost any location, including non-convective adiabats that are hot enough to explain observed features (such as the radius, etc.; see \autoref{sec:obs}). However, to do so, we needed to make a number of assumptions about both the deep radiative forcing as well as the 3D temperature structure. { In \autoref{sec:1D_1000b}, we show how we attempted to remove this second assumption by using 1D models of a sub-stellar-point day-side and a non-irradiated night-side to modify our 3D models radiative cooling profile. The aim of this being to constrain the pressure range at which we might expect advection to dominate over radiative effects and hence horizontally homogenise the deep atmosphere.}

\begin{figure*}[htbp!] %
\begin{centering}
\begin{subfigure}{0.49\textwidth}
\begin{centering}
\includegraphics[width=0.875\columnwidth]{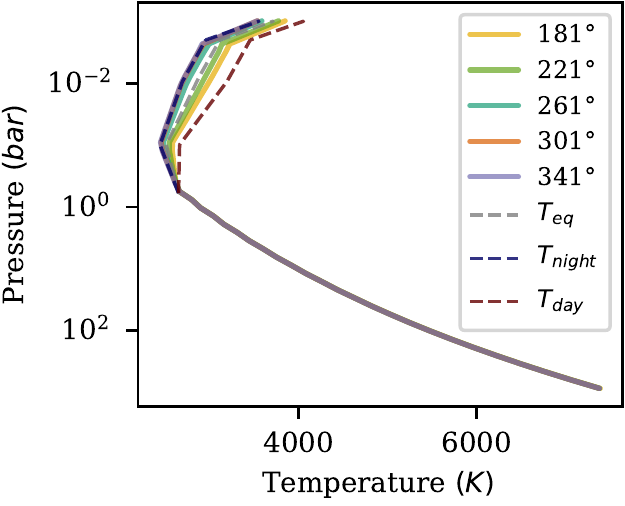}
\caption[]{Kepler-13Ab - 0.6 bar - 1D Match - Longitudinal Variations\label{fig:K13_T_P_0_5_bar_long} }
\end{centering}
\end{subfigure}
\begin{subfigure}{0.49\textwidth}
\begin{centering}
\includegraphics[width=0.875\columnwidth]{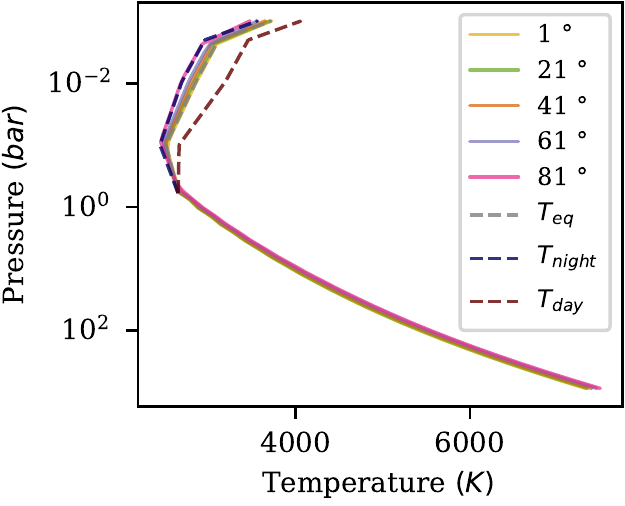}
\caption[]{Kepler-13Ab - 0.6 bar - 1D Match - Latitudinal Variations\label{fig:K13_T_P_0_5_bar_lat}  }
\end{centering}
\end{subfigure}
\begin{subfigure}{0.49\textwidth}
\begin{centering}
\includegraphics[width=0.875\columnwidth]{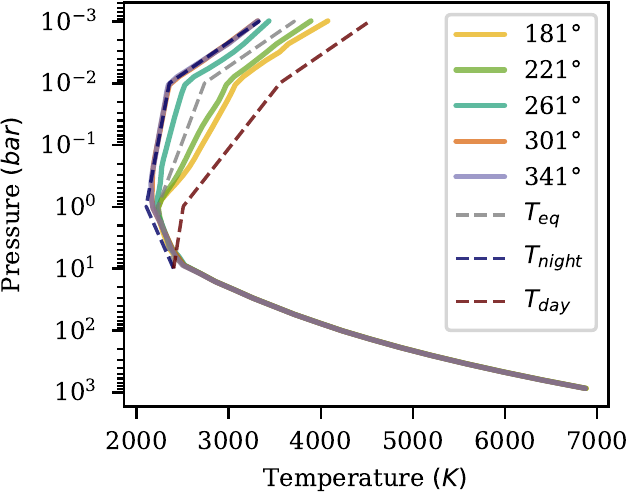}
\caption[]{KELT-1b - 10 bar - 1D Match - Longitudinal Variations\label{fig:Kelt1b_T_P_10_bar_long}  }
\end{centering}
\end{subfigure}
\begin{subfigure}{0.49\textwidth}
\begin{centering}
\includegraphics[width=0.875\columnwidth]{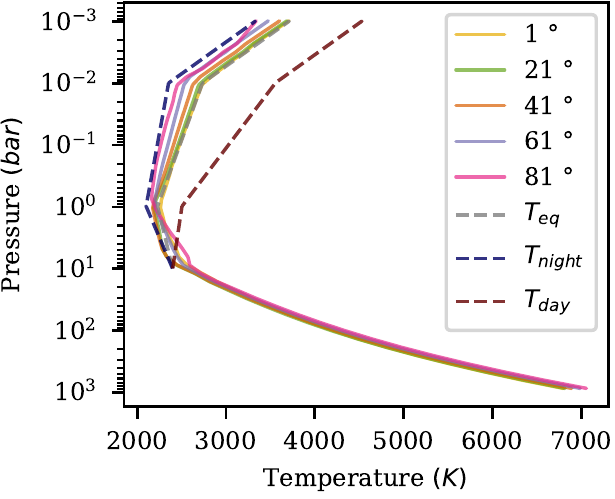}
\caption[]{KELT-1b - 10 bar - 1D Match - Latitudinal Variations\label{fig:Kelt1b_T_P_10_bar_lat}  }
\end{centering}
\end{subfigure}
\caption[]{ Snapshots of the latitudinally (left) or longitudinally (right) averaged T-P profile for two evolved models of Kepler-13Ab (top) and KELT-1b (bottom), both with a very shallow horizontal convergence pressure, at various longitudes or latitudes, respectively. { These models have been selected in order to match either the observed radius of the brown dwarf (KELT-1b), or the minimum deep atmosphere temperatures required to explain a water absorption feature in the brown dwarf spectra (Kepler-13Ab); } see \autoref{sec:obs} for more details.\label{fig:obs_PT_comp} }
\end{centering}
\end{figure*} 

\subsubsection{Freely varying the convergence pressure} \label{sec:free}
{ We start by exploring models of each brown dwarf with very loose limits on the location of the horizontal convergence pressure. More specifically, we simply require that the convergence pressure should be less than or equal to that required to sustain the corresponding, uninflated 1D models' convective adiabat, while also being high enough that it remains compatible with observations of the day-night temperature difference. \\}

{ To illustrate this, in \autoref{fig:KELT_T_P_comp}, we plot three exemplary models} of KELT-1b with horizontal convergence pressures of 1 bar (top), 10 bar (middle), and 40 bar (bottom). In all three models, the atmospheres are initialised with an adiabat based on the convergence pressure-temperature point given in \autoref{tab:equilbirum_temperature}. \\
For models of both KELT-1b and Kepler-13Ab, we find that the deep adiabat then starts to slowly, but significantly, heat. { However, the rate of heating slows as we move towards shallower (i.e. lower) horizontal convergence pressures. This results} in a smaller fractional increase in the deep adiabat's temperature, compared to the adiabat used to initialise it{: a $\sim3.7\%$ temperature increase at 1 bar, $\sim4.9\%$ at 10 bar, and $\sim6.1\%$ at 40 bar.}
This reduction in the deep heating rate can likely be linked to both the increase in thermal mass of the `deep' atmosphere, as well a reduction in the downward heating rate as the model's outer atmosphere becomes smaller and the equatorial jet becomes shallower. { The latter} can be attributed to the equatorial jet being primarily driven by day-night temperature differences.

This sensitivity of the deep heating rate to the depth of the convergence pressure is even more pronounced for SDSS1411B. While deeper convergence pressure models do show signs of very slight deep heating, said heating appears to slow much more quickly as we move towards shallower convergence pressures.
We explore how the deep heating rate changes with convergence pressure in more detail in \autoref{sec:flows}. 

Given this slightly tighter constraint on the horizontal convergence pressure (i.e. it must be deep enough for a significant, vertical advection driving, equatorial jet to develop), we next explored whether our models for Kepler-13Ab and KELT-1b can maintain{ advective adiabats that might explain the observed phenomena; for example, the observed water absorption feature of Kepler-13Ab or the inflated radius of KELT-1b.} 
We note that we excluded SDSS1411B from this part of the study since observations suggest that it should not be significantly inflated and, indeed, the observed radius is consistent with evolutionary models \citep{2014MNRAS.445.2106L}. 

For Kepler-13Ab, explaining the observed water absorption features (without invoking an additional thermal inversion) requires at minimum that the atmosphere is adiabatic for all pressures greater than around 0.6 bar. As for KELT-1b, in order to match the observed radius, comparisons with evolutionary and 1D atmospheric models { with altered (i.e. increased) internal temperatures} suggests that the internal adiabat should develop at pressures less than around 10 bar{. This is significantly shallower and, hence, hotter than the convective adiabat that develops at around 40 bar in our reference, radiative-convective, 1D model.} \\ 
In \autoref{fig:obs_PT_comp}, we plot (latitudinally to the left and longitudinally to the right) the averaged temperature-pressure profiles at various longitudes or latitudes, respectively, for the models of Kepler-13Ab (top) and KELT-1b (bottom). Here we find that when using a suitable (i.e. shallow) horizontal convergence pressure, both brown dwarf models are able to maintain an advective adiabat that agrees with the observational criteria{. That is to say} both longitudinally and, to a slightly lesser extent, latitudinally, the deep atmospheres have both converged onto an advective deep adiabat that is hot enough to explain the observed phenomena. We note that the difference in convergence rate between the longitudinal and latitudinal profiles can be linked to a combination of the relative strengths of zonal and meridional flows, and the amount of time that the simulations have been run for. More specifically, zonal dynamics tend to be much stronger than meridional dynamics thanks to the super rotating equatorial jet. As a result, after $\sim$200 Earth years of simulation time, the models have converged longitude wise but are still very slowly converging latitude wise. However, since the profiles are already close to being converged at all longitudes, we do not expect this ongoing slow latitudinal convergence to drastically impact the final temperature of the deep advective adiabat. \\

Even though  these deep advective adiabats are a good a match to the observations, two key questions remain before we can ascertain if this result is truly valid{. Firstly, we examine wether these are suitable pressures to use} when setting the day-night temperature difference to zero. Secondly, we consider what happens when we include radiative forcing in the deep atmosphere.
\begin{figure}[hbtp!] %
\begin{centering}
\includegraphics[width=0.95\columnwidth]{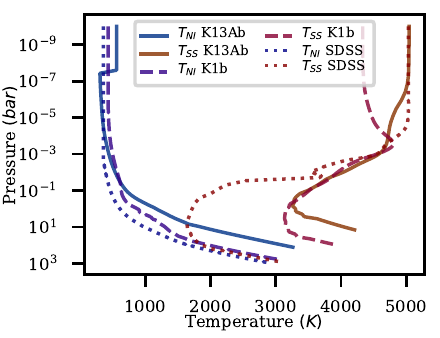}
\caption{ 1D models of sub-stellar `day-side' (reds) and non-irradiated `night-side' (blues) for the three brown dwarfs under consideration here: Profiles for Kepler-13Ab are plotted with solid lines, with dashed lines used for KELT-1b, and dotted lines for SDSS1411B. \label{fig:1D_1000b} }
\end{centering}
\end{figure}
\begin{figure*}[htbp!] %
\begin{centering}
\begin{subfigure}{0.45\textwidth}
\begin{centering}
\includegraphics[width=0.99\columnwidth]{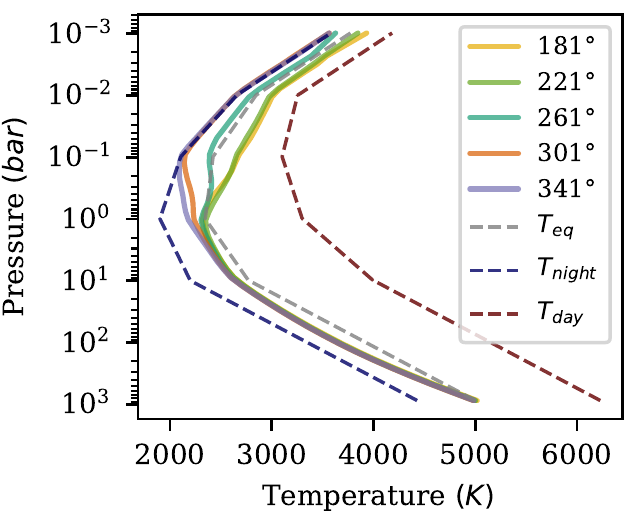}
\caption[]{Kepler-13Ab - Longitudinal variations \label{fig:1000b_K13_multilong} }
\end{centering}
\end{subfigure}
\begin{subfigure}{0.45\textwidth}
\begin{centering}
\includegraphics[width=0.99\columnwidth]{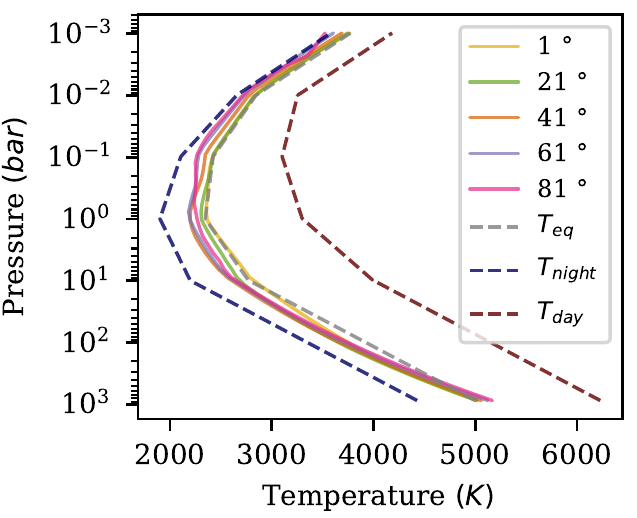}
\caption[]{Kepler-13Ab - Latitudinal variations \label{fig:1000b_K13_multilat} }
\end{centering}
\end{subfigure}
\begin{subfigure}{0.45\textwidth}
\begin{centering}
\includegraphics[width=0.99\columnwidth]{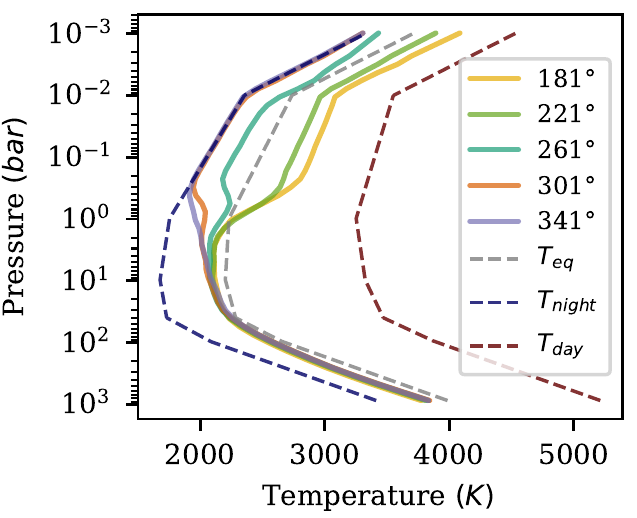}
\caption[]{KELT-1b - Longitudinal variations \label{fig:1000b_KELT1b_multilong} }
\end{centering}
\end{subfigure}
\begin{subfigure}{0.45\textwidth}
\begin{centering}
\includegraphics[width=0.99\columnwidth]{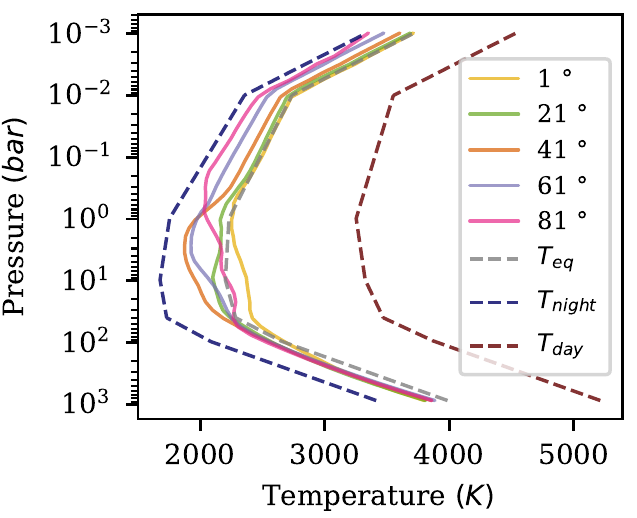}
\caption[]{KELT-1b - Latitudinal variations   \label{fig:1000b_KELT1b_multilat} }
\end{centering}
\end{subfigure}
\caption[]{Snapshots of the latitudinally (left) or longitudinally (right) averaged T-P profile for two evolved models of Kepler-13Ab (top) and KELT-1b (bottom) at various longitudes or latitudes respectively. These models include radiative forcing, via Newtonian cooling, at all pressures. This cooling is based upon a combination of the equilibrium 1D radiative timescale profile (grey) { including the} observed day-night temperature difference in the outer atmosphere, and 1D sub-stellar and non-irradiated 1D models in the deep atmosphere{. The overall result is a combined }Newtonian cooling model with a day-side (at the sub-stellar point) forcing profile that corresponds to the red dashed line and a night-side forcing profile that corresponds to the blue dashed line. \label{fig:1000b_multilong_lat}}
\end{centering}
\end{figure*} 

\subsubsection{Using 1D day-night models to constrain the convergence pressure} \label{sec:1D_1000b}
In order to better constrain the horizontal convergence pressures in our brown dwarf models, we next modify our models to remove any assumptions about where the day-night temperature difference goes to zero. \\

{ To do this, we must go beyond using just the equilibrium 1D models and instead look at day-side and night-side profiles explicitly. To that end, we next}  use a sub-stellar point profile to represent the day-side { of our model} and a non-irradiated planet to represent the cold night-side. We plot these 1D model profiles for all three of the brown dwarfs under consideration here in \autoref{fig:1D_1000b}. Here, the sub-stellar day-side profiles are shown in shades of red and the non-irradiated night-side profiles in shades of blue. Further each brown dwarfs profiles uses a different line-style: Solid for Kepler-13Ab, dashed for KELT-1b, and dotted for SDSS1411B.

Immediately these 1D profiles reveal a stark difference between Kepler-13Ab and KELT-1b, and SDSS1411B. For the former two brown dwarfs we find that the 1D models do not converge at any pressure. Whereas for SDSS1411B, we find that the sub-stellar day and non-irradiated night 1D models converge at $\sim$100 bar. Since we do not expect SDSS1411B to significantly inflate, and hence we do not expect to find significant differences between 1D models and observations, we set 100 bar as the preliminary horizontal convergence pressure for models of SDSS1411B (see \autoref{sec:radiative_stability} for results).

As for Kepler-13Ab and KELT-1b, at first glance it might appear that we can simply use these 1D profiles to set the day-side and night-side Newtonian Cooling forcing profiles in our models and then evolve said models to a steady state. However, doing this would be not only a failure to understand the distinction between the local nature of the (i.e. sub-stellar or non-irradiated) 1D models and the equilibrium nature of Newtonian cooling, but also the fundamental meaning of the term radius inflation: an increase in the observed radius that cannot be explained by 1D models alone. 
The 1D models lack horizontal advective mixing{, such as the super-rotating jet which advectively cools the day-side and heats the night-side. As a result,} when we look at the temperature difference between the 1D sub-stellar day and non-irradiated night profiles in the outer atmosphere, we find a day-night contrast that is significantly higher than what has been observed. 
On the other hand, equilibrium Newtonian cooling requires that (as the name might imply) the relaxation temperature profiles represent the equilibrium temperature state. Thus, if we use the non-advective and non-redistributive  T-P profiles from the 1D models with our Newtonian cooling scheme, while we might find that our models develop the required super-rotating jet, any advective cooling or heating on the day or night sides would be eliminated, as the relaxation forces the atmosphere back towards the imposed, non-equilibrium, profiles. \\
In summary, the use of Newtonian cooling means that our outer atmosphere temperature profile must be prescriptive rather than descriptive. We note that, of course, using a GCM that includes a full description of radiative effects would eliminate this problem. However, the vast increase in computational cost means that such a model was not suitable for this initial exploration of the deep atmospheres of brown dwarfs (see \autoref{sec:conclusion} for a discussion of how next-generation radiative GCMs might be put to work to explore deep brown dwarf atmospheres in future studies).

However, the above does not mean that these 1D sub-stellar day and non-irradiated night profiles are not still useful for constraining the pressure at which the atmosphere horizontally converges. To that end, we  ran a pair of models of Kepler-13Ab and KELT-1b with a{ Newtonian cooling} profile based on the sub-stellar day and non-irradiated night 1D models in the deep atmosphere, and the equilibrium 1D model { modified by} the observed day-night temperature difference in the outer atmosphere. 
{ This combined forcing profile allows us to investigate how the deep atmosphere (and, in particular, the advective dynamics) responds to the presence of a (somewhat) physically motivated deep day-night temperature difference. It also ensures that the outer atmosphere dynamics and, hence, the resulting advective downflows, are appropriate for both comparisons with observations and with our ‘core’ equilibrium forcing models.} \\
The resulting combined day (red dashed), night (blue dashed) and `equilibrium' (grey dashed) temperature forcing profiles can be seen for both Kepler-13Ab (top) and KELT-1b (bottom) as part of \autoref{fig:1000b_multilong_lat}. { In this figure, we plot the latitudinally (left) and longitudinally (right) averaged T-P profiles at various longitudes or latitude, respectively, for both brown dwarf models.} 
Here, we see that both brown dwarf models show significant signs of horizontal convergence, particularly longitudinally{. However latitudinally, the models are still slowly converging, an effect which can once} again be linked to the differences in strength between zonal and meridional flows. \\
An analysis of the longitudinal variation profiles suggests that Kepler-13Ab starts to converge between about 1 and { $\sim$4-5 bar}, and KELT-1b starts to converge between about 10 and {$\sim$40-50 bar}. The latitudinal variation profiles, at the current simulation time of approximately 300 Earth years, tend to support this, particularly for KELT-1b, where we start to see some convergence around 10 bar, which gets stronger as we shift deeper towards 40 bar. \\

{ As discussed and shown above}, we have explored a number of models with convergence pressures close to or within this range and shown that a hot, deep adiabat is able to develop. However, one question remains as to whether these advective adiabats are stable when we introduce deep radiative forcing back into the system. We explore this question in the following section. 

\subsection{Sensitivity of deep advective adiabats to the inclusion of deep radiative forcing} \label{sec:radiative_stability}
\begin{figure*}[htbp!] %
\begin{centering}
\begin{subfigure}{0.4\textwidth}
\begin{centering}
\includegraphics[width=0.99\columnwidth]{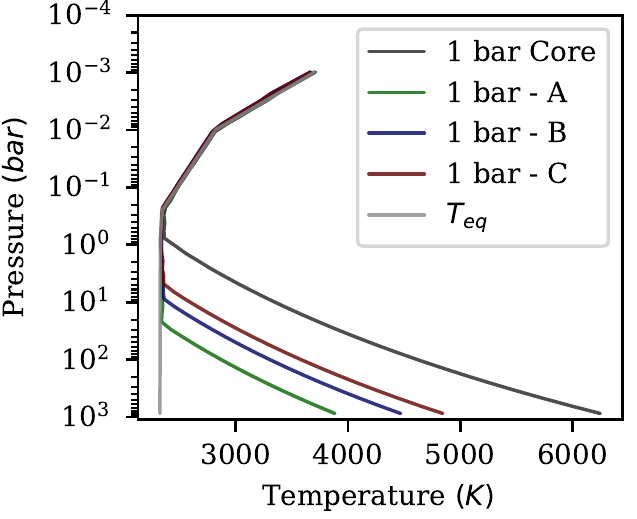}
\caption[]{Kepler-13Ab - Temperature-pressure profiles \label{fig:Kepler-13Ab_shallower_rf_TP} }
\end{centering}
\end{subfigure}
\begin{subfigure}{0.4\textwidth}
\begin{centering}
\includegraphics[width=0.99\columnwidth]{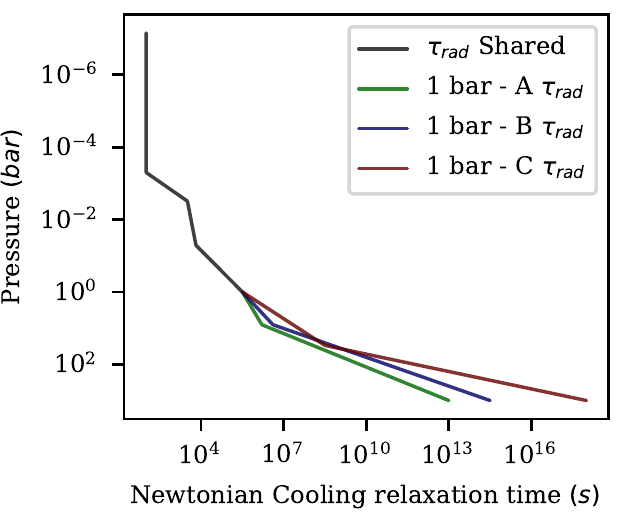}
\caption[]{Kepler-13Ab - Radiative forcing timescales  \label{fig:Kepler-13Ab_shallower_rf} }
\end{centering}
\end{subfigure}
\begin{subfigure}{0.4\textwidth}
\begin{centering}
\includegraphics[width=0.99\columnwidth]{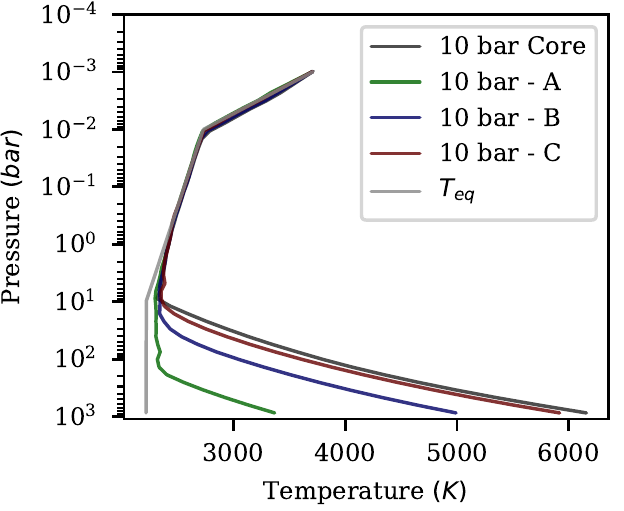}
\caption[]{KELT-1b - Temperature-pressure profiles \label{fig:Kelt1b_shallower_rf_TP} }
\end{centering}
\end{subfigure}
\begin{subfigure}{0.4\textwidth}
\begin{centering}
\includegraphics[width=0.99\columnwidth]{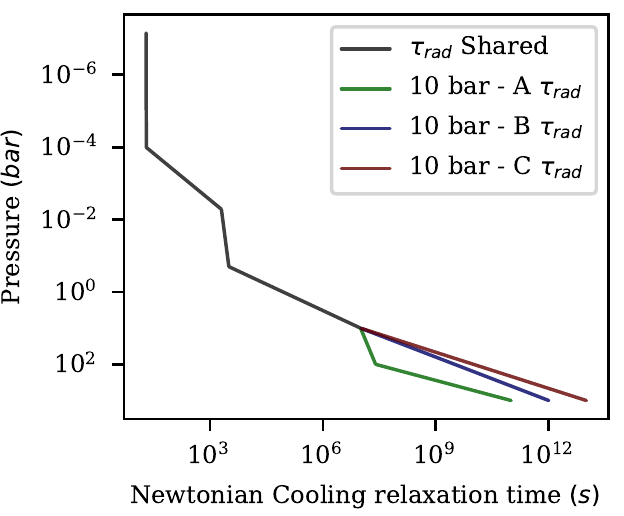}
\caption[]{KELT-1b - Radiative forcing timescales   \label{fig:Kelt1b_shallower_rf} }
\end{centering}
\end{subfigure}
\caption[]{ Snapshots of the T-P profile (left) for a series of test models investigating the effect of deep radiative forcing, via deep isothermal Newtonian cooling, on already evolved (dark grey) models of both Kepler-13Ab (top) and KELT-1b (bottom) with a shallower horizontal convergence pressure. For each brown dwarf, we explore models with increasingly weak deep forcing, as shown by the matching profile on the right (where the full Newtonian cooling timescale profile is found by combining the dark grey profile at lower pressures and the matching coloured profile at higher pressures). \label{fig:K13_KELT_shallow_DF}  }
\end{centering}
\end{figure*}
\begin{figure*}[htbp!] %
\begin{centering}
\begin{subfigure}{0.4\textwidth}
\begin{centering}
\includegraphics[width=0.99\columnwidth]{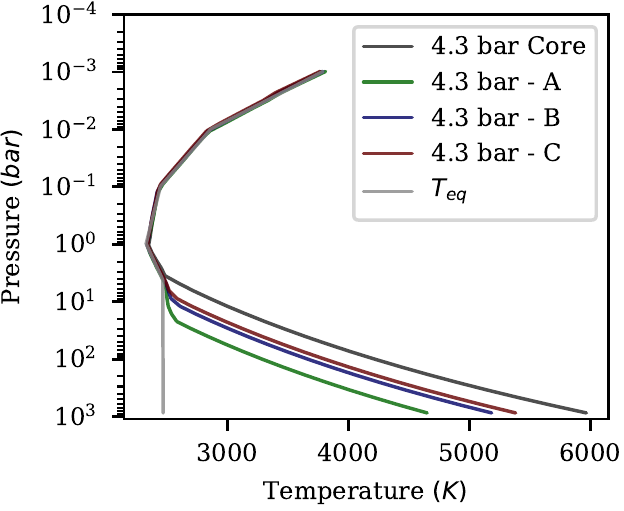}
\caption[]{Kepler-13Ab - Temperature-pressure profiles \label{fig:Kepler-13Ab_deep_rf_TP} }
\end{centering}
\end{subfigure}
\begin{subfigure}{0.4\textwidth}
\begin{centering}
\includegraphics[width=0.99\columnwidth]{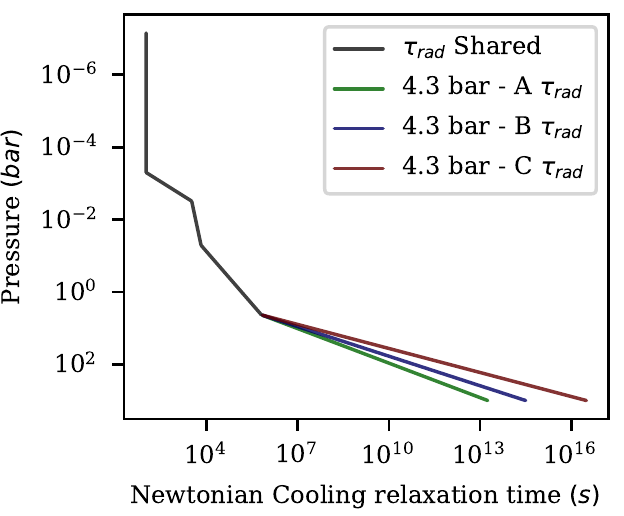}
\caption[]{Kepler-13Ab - Radiative forcing timescales  \label{fig:Kepler-13Ab_deep_rf} }
\end{centering}
\end{subfigure}
\begin{subfigure}{0.4\textwidth}
\begin{centering}
\includegraphics[width=0.99\columnwidth]{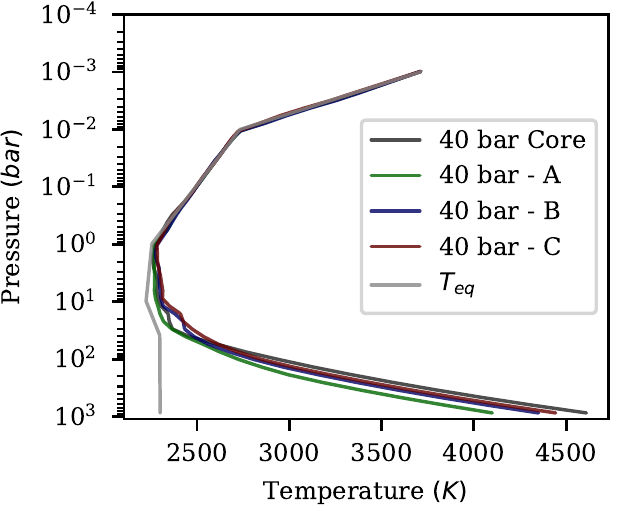}
\caption[]{KELT-1b - Temperature-pressure profiles \label{fig:Kelt1b_deep_rf_TP} }
\end{centering}
\end{subfigure}
\begin{subfigure}{0.4\textwidth}
\begin{centering}
\includegraphics[width=0.99\columnwidth]{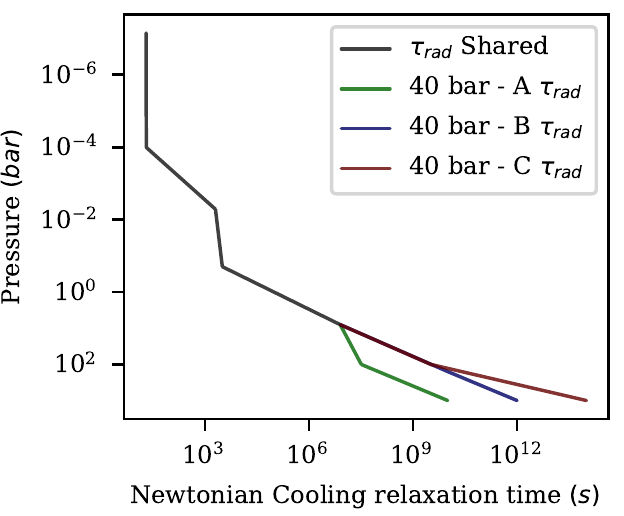}
\caption[]{KELT-1b - Radiative forcing timescales   \label{fig:Kelt1b_deep_rf} }
\end{centering}
\end{subfigure}
\caption[]{ Snapshots of the T-P profile (left) for a series of test models investigating the effect of deep radiative forcing, via deep isothermal Newtonian cooling, on already evolved (dark grey) models of both Kepler-13Ab (top) and KELT-1b (bottom) with a deeper horizontal convergence pressure. For each brown dwarf, we explored models with increasingly weak deep forcing, as shown by the matching profile on the right, where the full Newtonian cooling timescale profile is found by combining the dark grey profile at lower pressures and the matching coloured profile at higher pressures.  \label{fig:K13_KELT_deep_DF} } 
\end{centering}
\end{figure*}
\begin{figure}[htbp!] %
\begin{centering}
\includegraphics[width=0.99\columnwidth]{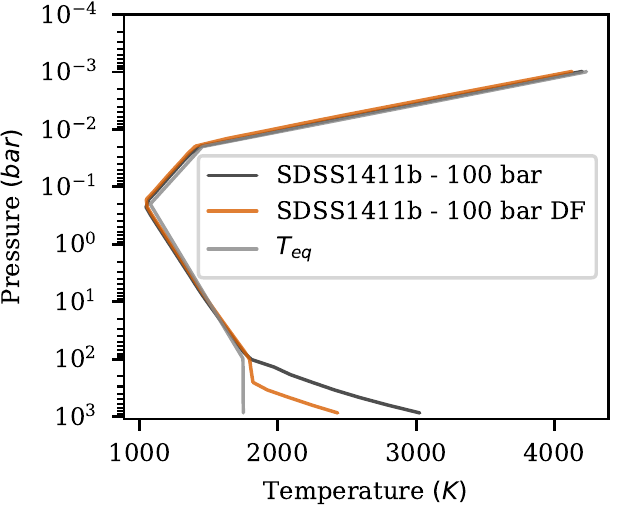}
\caption[]{Snapshot of the still-cooling T-P profile for a variant of the 100 bar convergence pressure model of SDSS1411B (dark grey) that includes radiative forcing, via Newtonian cooling, at all pressures { (orange - deep forcing, DF)}. This forcing is towards an equilibrium profile in the outer atmosphere ($P<P_{converge}$) and towards an isotherm in the deep atmosphere ($P>P_{converge}$). We note that we only include a single model with deep radiative forcing here since models with even weaker deep cooling behave similarly, just with a different evolution and cooling timescale.  \label{fig:SDSS_DF} } 
\end{centering}
\end{figure}

At this point, we have a tighter constraint on the pressure at which we might expect the deep atmospheres of Kepler-13Ab, KELT-1b, and SDSS1411B to horizontally converge{.  In addition, we have confirmed that for models with no deep radiative forcing, all three brown dwarfs can maintain a deep advective adiabat at these convergence pressures{. We next} explore how stable these advective adiabats are when radiative effects { are no longer limited to the outer atmosphere}. 

{ To achieve this, we} took already evolved models of all three brown dwarfs and explored their evolution when we introduced various strengths of deep radiative forcing (controlled via the radiative forcing timescale) towards an isotherm that has been set to the same temperature as that found at the horizontal convergence pressure. We note that the analysis below is split into two sections representing the two observed dynamical regimes: Kepler-13AB and KELT-1b, which both behave similarly to HD209458b and show significant signs of deep heating, and SDSS1411B, which appears to struggle to heat the deep atmosphere significantly above the initial convective adiabat from the 1D models, even when radiative { effects are limited to the outer atmosphere}.\\

\subsubsection{Kepler-13Ab and KELT-1b}
We start by exploring models { of Kepler-13Ab and KELT-1b} with horizontal convergence pressures that fall towards the shallower end of { the calculated convergence pressures (\autoref{sec:1000bar})}: 1 bar for Kepler-13Ab and 10 bar for KELT-1b.

\autoref{fig:K13_KELT_shallow_DF} shows the T-P profiles (left) after $\sim$250 Earth years of simulation time for three models of both Kepler-13Ab (top) and KELT-1b (bottom), each with increasingly long radiative cooling timescales (right).
{ For all deep radiative forcing models shown here, the base model used for initialisation is shown in dark grey and the temperature profile they are being cooled or forced towards is shown in light grey.} \\
Here, we find that, for both brown dwarfs, the models with the strongest deep forcing (labelled A in \autoref{fig:K13_KELT_shallow_DF}){ that most closely matches the radiative timescales from the 1D radiative-equilibrium models} show significant signs of ongoing deep cooling towards the forced isotherm. \\
For Kepler-13Ab, as we increase the radiative timescale of the deep forcing (B and C), we find a corresponding decrease in the rate of deep cooling{. This effect is clearest in the weakest deep forcing model (C), which is cooling} at an increasingly slow rate that suggests that it is asymptotically approaching a steady state in which at least some form of an advective adiabat is maintained. 
However, this stable deep adiabat is significantly cooler than either the initialisation profile or the convective adiabat that develops in the 1D equilibrium models (or which would develop in these models if we included the physics required for convective motions){.  This raises} some doubts as to whether 1 bar is the correct horizontal convergence pressure for Kepler-13Ab.   \\
As for KELT-1b, due to differences in the radiative timescale parametrisation between the fastest (A) deep forcing model and the slower deep forcing models (B and C -  which do not include the 1D models region of near constant $\tau_{rad}$ between 10 and 100 bar), we find that the slower deep forcing models maintain steady-state advective adiabats that are much hotter than found in the strong deep forcing case (A).  
In particular, the slowest deep forcing model (C) reveals a steady advective adiabat that is only slightly cooler than that of the base model, and thus is both close to the temperature that would be required to explain the radius inflation of KELT-1b (see \autoref{fig:obs_PT_comp}), { as well as being notably hotter than the 1D convective adiabat}.\\

{ This apparent sensitivity of models with shallower horizontal convergence pressures to deep radiative forcing raises some questions about the suitability of such convergence pressures. 
As such, w}e next explore models of Kepler-13Ab and KELT-1b with horizontal convergence pressures that fall towards the deeper end of { the calculated convergence pressure range (\autoref{sec:1000bar})}: 4.3 bar for Kepler-13Ab and 40 bar for KELT-1b. 

\autoref{fig:K13_KELT_deep_DF} shows the T-P profiles (left) after $\sim$300 Earth years of simulation time for three models of both Kepler-13Ab (top) and KELT-1b (bottom) with increasingly slow deep radiative forcing towards an isotherm (right). However, unlike in the models shown in \autoref{fig:K13_KELT_shallow_DF}, none of the deep forcing models shown here exhibit signs of ongoing cooling. \\
{ Starting with Kepler-13Ab, we find that all models show at least some signs of cooling from the initial adiabat. However this cooling is limited for models with longer radiative timescales (B and C), with both maintaining steady-state advective adiabats that are at least slightly hotter than the 1D models' convective adiabat. }
Only the very strongest deep forcing model (A) shows signs of a significant cooling of the deep adiabat, and even then it remains hotter than the weaker deep forcing model (B) with a 1 bar convergence pressure. This apparent slowdown in the deep cooling rate is likely linked to differences in vertical heat transport. { We explore these differences in more} detail in \autoref{sec:flows}. \\
Moving onto KELT-1b, { we find that the differences between models with and without deep radiative forcing have become even smaller. The models with slower deep forcing (B and C) maintain steady-state advective adiabats that are at least slightly hotter than the 1D models' convective adiabat. And as for the model with stronger forcing (A - which includes the near constant radiative forcing timescale between approximately 10 and 100 bar), here we also find a steady advective adiabat that is only slightly cooler (by $\sim$200K at 300 bar - around a 6\% difference) than the 1D model.} \\ 

These results suggest that models of Kepler-13Ab or KELT-1b with a 4.3 bar or 40 bar convergence pressure, respectively, should be able to maintain a stable advective adiabat that is hotter than the convective adiabat that a 1D equilibrium model would suggest{. That is to say they} exhibit at least some form of radius inflation. We note that while all the models above exhibit at least some level of cooling from the initial state, low levels of cooling are not unexpected. Not only was a similar effect noted by \citetalias{2019A&A...632A.114S} for models of HD209458b with low levels of deep radiative forcing, but this forcing was designed to be somewhat exaggerated in order to emphasise its effect. After all we do not expect the true equilibrium state of the deep atmosphere to be an isotherm. \\
However, this does not mean that the trend revealed by the deep forcing is not real{, as even when we switch to forcing} the deep atmosphere towards the 1D models' convective adiabat, the models with a deeper convergence pressure are much more stable to the effects of deep radiative dynamics, and the models that are significantly impacted by deep radiative forcing remain so. We explore why this might be the case in \autoref{sec:flows}. \\
\subsubsection{SDSS1411B}
Moving on to SDSS1411B, \autoref{fig:SDSS_DF} plots T-P profiles for two 100 bar convergence pressure model { atmospheres.  We show the base }model in dark grey, which, after $\sim$70 Earth years of simulation time has only very slightly warmed from its initial adiabat, and a deep forcing model (orange) which is still steadily cooling (at a near constant rate) after an additional $\sim30$ Earth years of simulation time.
{ We note that we only included a single model of SDSS1411B with deep radiative forcing due to similarities in the results. Unless we set the deep radiative timescale to be drastically slower than the 1D models predict, for instance, $\tau_{rad} = 10^{18\rightarrow20}$ at 1000 bar, all deep forcing models show signs of ongoing deep cooling, albeit at steadily slower rates as the deep radiative timescale is lengthened.\\ }
 We discuss why this might be the case in \autoref{sec:flows}. However, briefly, it appears that downward advective transport is much weaker in SDSS1411B atmosphere than in the atmospheres of Kepler-13Ab and, KELT-1b. This has the effect of making any advective adiabat that develops in SDSS1411B much more sensitive to deep radiative effects.}

These results suggest that (as observations require) the formation of a stable hot and deep adiabat is unlikely to lead to any significant inflation of SDSS1411B atmosphere{. This can be understood as occurring} because not only does advection struggle to heat the deep atmosphere significantly, but even weak radiative effects cool the advective adiabat below the convective adiabat 1D models suggest would develop if we included convective physics in our models. 
\subsection{Differences in flows and circulations between Kepler-13Ab/KELT-1b and SDSS1411B} \label{sec:flows}
\begin{figure*}[htbp!] %
\begin{centering}
\begin{subfigure}{0.45\textwidth}
\begin{centering}
\includegraphics[width=0.99\columnwidth]{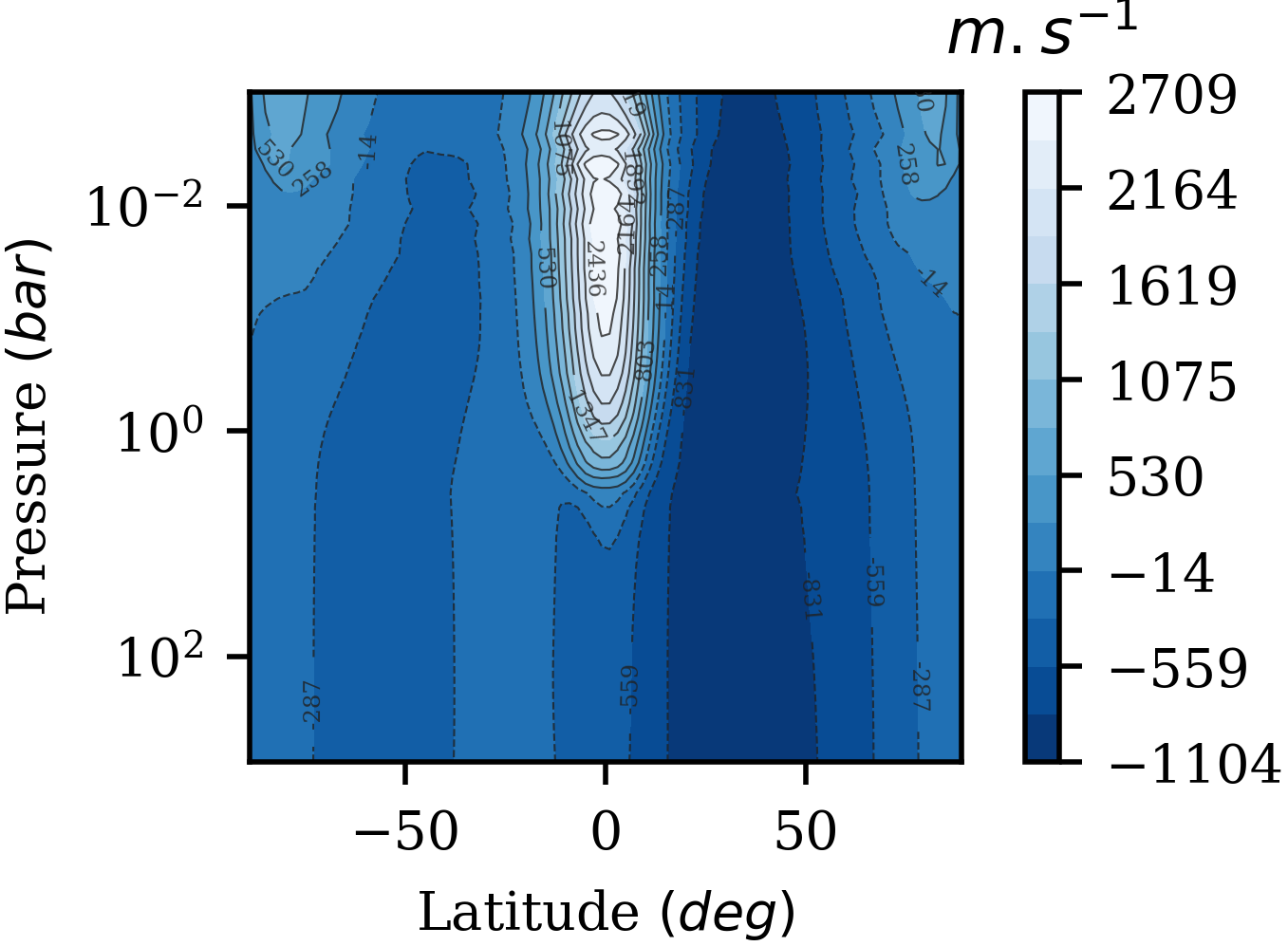}
\caption[]{Kepler-13Ab - Zonal Wind \label{fig:Kepler-13Ab_zonal_wind} }
\end{centering}
\end{subfigure}
\begin{subfigure}{0.45\textwidth}
\begin{centering}
\includegraphics[width=0.99\columnwidth]{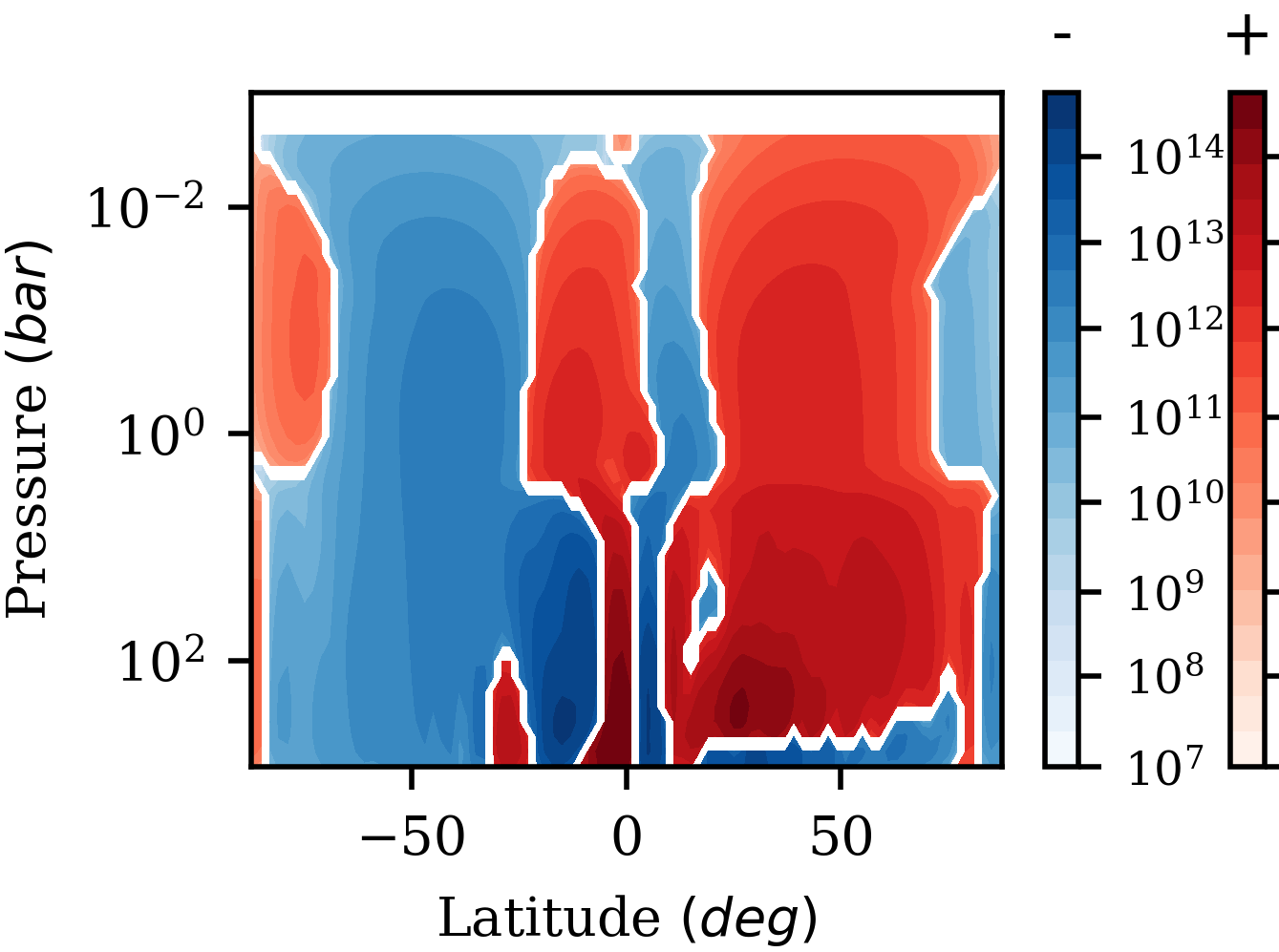}
\caption[]{Kepler-13Ab - Meridional Circulation \label{fig:Kepler-13Ab_SF} }
\end{centering}
\end{subfigure}
\begin{subfigure}{0.45\textwidth}
\begin{centering}
\includegraphics[width=0.99\columnwidth]{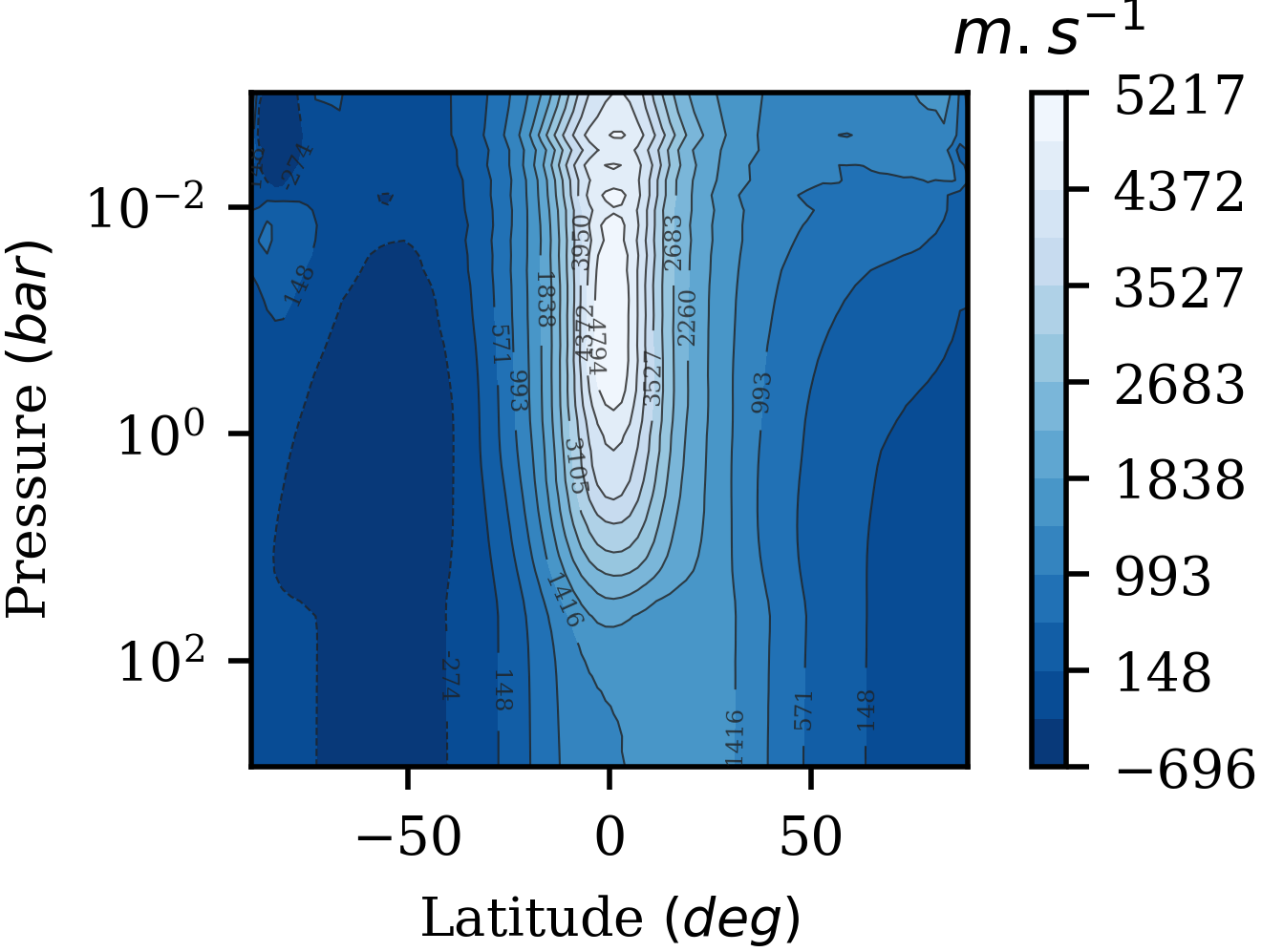}
\caption[]{KELT-1b - Zonal Wind \label{fig:Kelt1b_zonal_wind} }
\end{centering}
\end{subfigure}
\begin{subfigure}{0.45\textwidth}
\begin{centering}
\includegraphics[width=0.99\columnwidth]{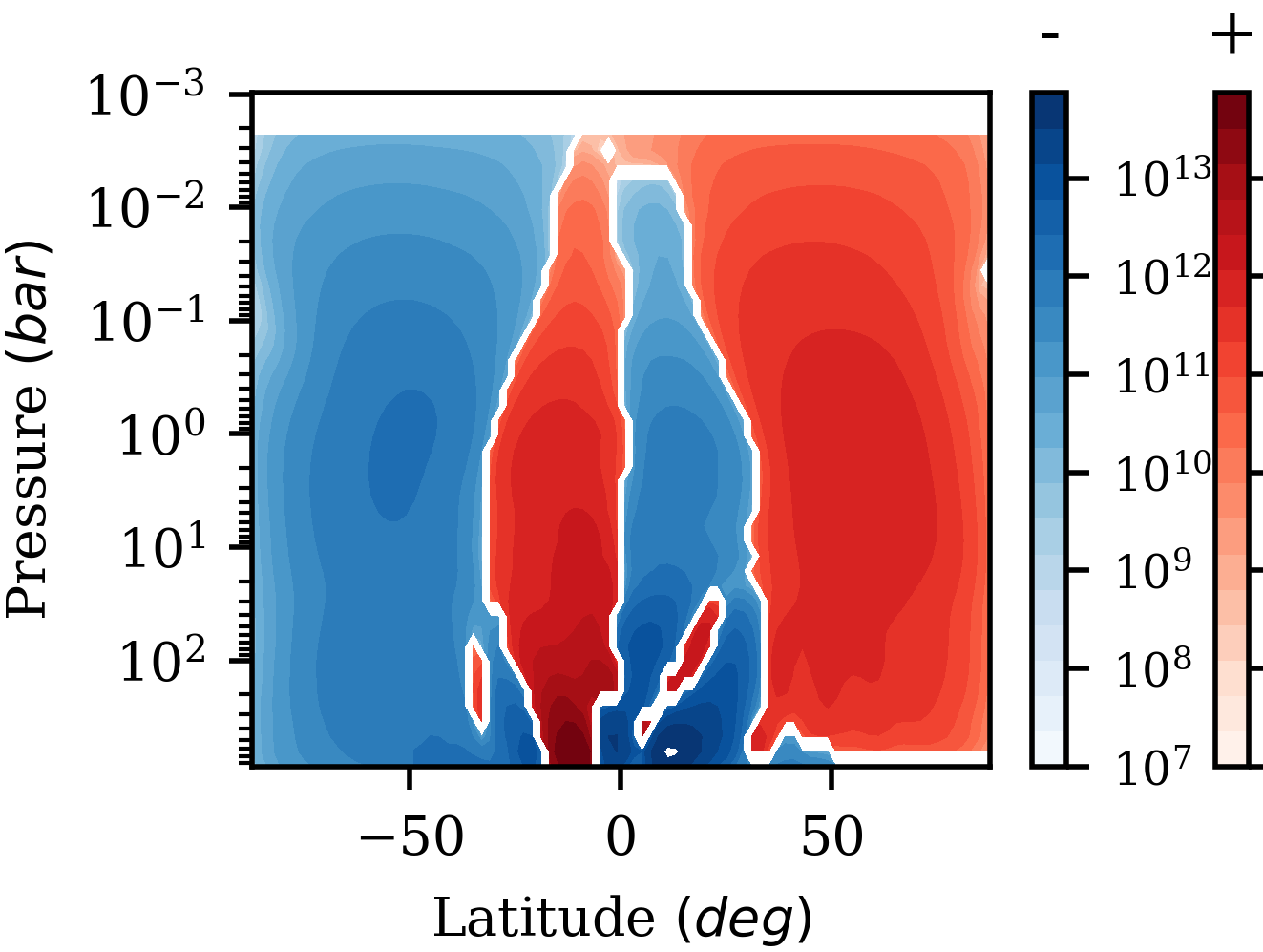}
\caption[]{KELT-1b - Meridional Circulation \label{fig:Kelt1b_SF} }
\end{centering}
\end{subfigure}
\begin{subfigure}{0.45\textwidth}
\begin{centering}
\includegraphics[width=0.99\columnwidth]{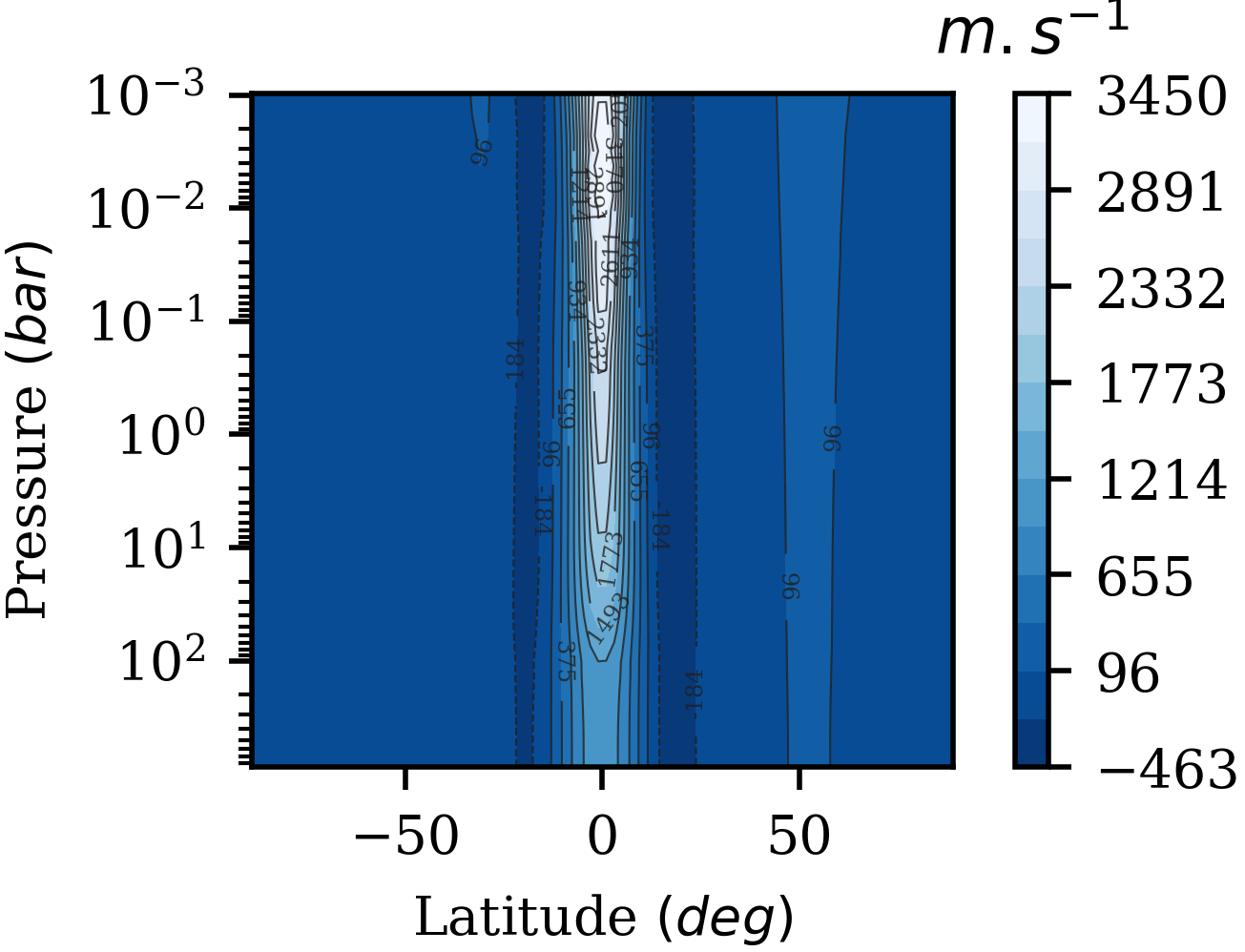}
\caption[]{SDSS1411B - Zonal Wind   \label{fig:SDSS_zonal_wind} }
\end{centering}
\end{subfigure} 
\begin{subfigure}{0.45\textwidth}
\begin{centering}
\includegraphics[width=0.99\columnwidth]{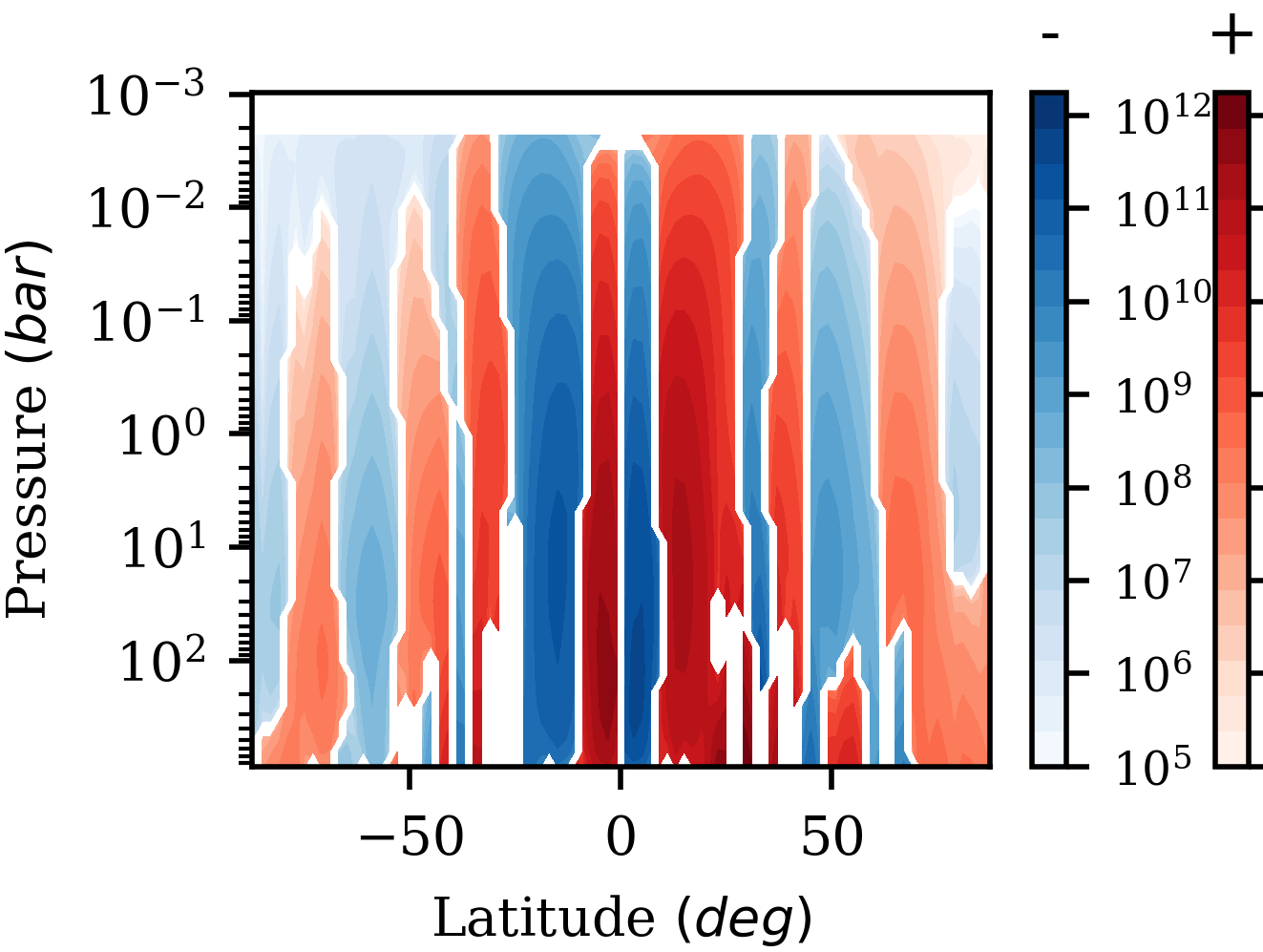}
\caption[]{SDSS1411B - Meridional Circulation   \label{fig:SDSS_SF} }
\end{centering}
\end{subfigure}
\caption[]{ Zonally and temporally averaged (over a period of $\sim$ 3 years) zonal wind profiles (left) and meridional circulation streamfunctions (right) for Kepler-13Ab (top), KELT-1b (middle), and SDSS1411B (bottom). In the zonal wind profiles easterly winds are positive and westerly winds are negative, and in the meridional circulation profile, we plot the stream function using a logged scale in order to clearly illustrate the full circulation profile{. Here, }clockwise circulations are shown in red and anti-clockwise in blue - these circulations combine in all models to reveal an equatorial downflow at all pressures.  \label{fig:flows_and_dynamics} }
\end{centering}
\end{figure*}
\begin{figure*}[htbp!] %
\begin{centering}
\begin{subfigure}{0.49\textwidth}
\begin{centering}
\includegraphics[width=0.95\columnwidth]{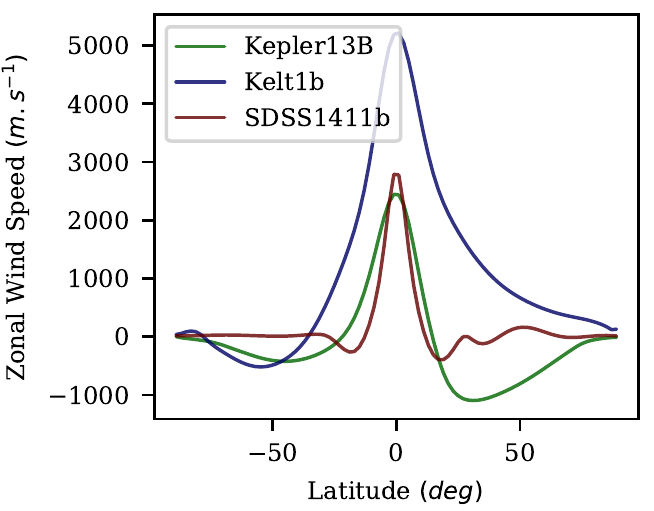}
\caption[]{0.1 bar \label{fig:zonal_wind_profile_0_1} }
\end{centering}
\end{subfigure}
\begin{subfigure}{0.49\textwidth}
\begin{centering}
\includegraphics[width=0.95\columnwidth]{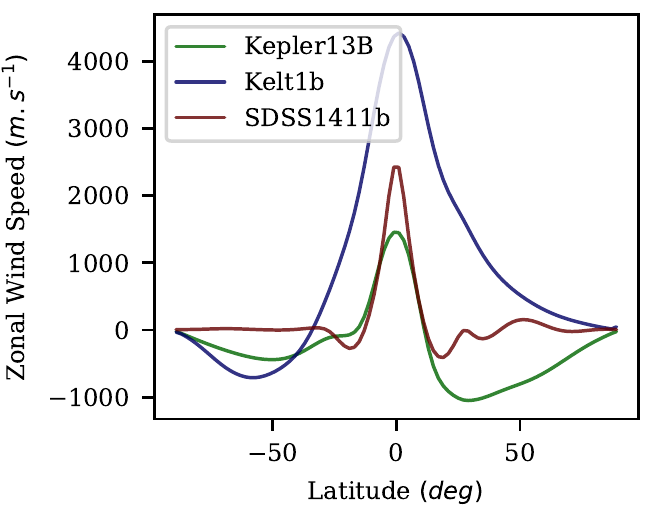}
\caption[]{1 bar  \label{fig:zonal_wind_profile_1}  }
\end{centering}
\end{subfigure}
\caption[]{Temporally and zonally averaged, over a period of $\sim$ 3 years, profiles of the zonal wind at 0.1 and 1 bar for deep horizontal convergence pressure models of Kepler-13Ab (green), KELT-1b (blue), and SDSS1411B (red). Here, easterly zonal winds are positive and westerly zonal winds are negative. \label{fig:zonal_wind_profiles} } 
\end{centering}
\end{figure*} 
{ In \autoref{sec:radiative_stability} we introduced two key results: 1) models of Kepler-13Ab and KELT-1b with a deeper} convergence pressure (4.3 and 40 bar respectively) can maintain a hot advective adiabat even in the presence of low levels of deep radiative forcing{. At the same time, models of SDSS1411B with a 100 bar convergence pressure struggle to maintain a hot advective adiabat in the presence of deep radiative forcing unless said }deep radiative effects are very weak (and even then the deep advective adiabat is not significantly warmed compared to the 1D models' convective adiabat); 2) models of Kepler-13Ab and KELT-1b with a shallower convergence pressure (i.e. 1 and 10 bar, respectively) appear to have a have harder time maintaining a hot advective adiabat in the presence of deep radiative effects than models with a deeper convergence pressure. Explanations for both of these results can be found by looking at both the flows and circulations that drive the vertical advective heating responsible for the formation of a hot and deep advective adiabat, as well as the associated energy fluxes. \\

\subsubsection{Zonal flows and meridional circulations}
{ We start, in \autoref{fig:flows_and_dynamics}, by exploring the zonally and temporally averaged zonal wind profiles (left) and meridional circulation profiles (right) for models of Kepler-13Ab (4.3 bar - top), KELT-1b (40 bar - middle), and SDSS1411B (100 bar - bottom) with deep convergence pressures. We note that the meridional circulation profile is also know as the streamfunction of the mass-flux on the meridional (i.e. latitudinal-pressure) plane.}

Here, we once again find that our models fall into two distinct regimes:
Kepler-13Ab and KELT-1b both exhibit zonal wind profiles that are dominated by a broad equatorial super-rotating jet that extends down to around the convergence pressure, and which tends to be flanked by much weaker, and even broader, higher latitude counterflows. Whereas SDSS1411B exhibits a very narrow jet that extends to the bottom of the atmosphere and which is flanked by similarly narrow, weaker, counterflows (which themselves give way to even weaker easterly flows etc.){.  We note that }very similar flow structures for very short period brown dwarfs orbiting white dwarfs were found by \citet{2020MNRAS.496.4674L} and \citet{2020ApJ...902...27T}. 

This difference in the latitudinal extent of both the equatorial jets, as well as the the flanking counterflows, can more easily be seen in \autoref{fig:zonal_wind_profiles}, which plots slices of the zonally and temporally averaged zonal wind at 0.1 bar (left) and 1 bar (right). Here, we can clearly see that at lower pressures, both the super-rotating easterly jet and the flanking westerly counter-flows of SDSS1411B are significantly narrower than the equivalent flows for either Kepler-13Ab or KELT-1b. Furthermore, we find that the equatorial jet that develops in the atmosphere of Kepler-13Ab is clearly weakening and narrowing as we get closer to the horizontal convergence pressure (and hence the `removal' of the day-night temperature difference that drives the jet){. A }similar effect was found for the jet of KELT-1b as we approach 40 bar, and to a lesser extent for the narrow jet of SDSS1411B as we approach 100 bar.

Evidence for the dynamical differences between our brown dwarf models can also be seen in the meridional circulation profiles (\autoref{fig:flows_and_dynamics}, right side). As might be expected, due to the super-rotating jet being the dominating driving force behind the flows and circulations in highly irradiated, tidally locked, planets, the meridional circulations profiles once again fall into two distinct regimes:
the circulation profiles of both Kepler-13Ab and KELT-1b are very reminiscent of those found for HD209458b (e.g. \citetalias{2019A&A...632A.114S}){.  We find }a clear equatorial downflow, slightly higher-latitude upflows, and weaker polar (or as the case is for Kepler-13Ab, near polar) downflows{. These }flows can all be linked back to the super-rotating equatorial zonal jet. \\
For example, in the meridional circulation profile of Kepler-13Ab we can see a clear difference between the near equator circulations in the outer atmosphere and in the deep atmosphere. In the outer atmosphere, the circulations cells extend horizontally to around the same latitude as the equatorial jet, whereas in the deep atmosphere the circulations become much narrower as rather than being driven flows associated with the zonal winds, they are being driven by the adiabatic redistribution of thermal energy that the outer atmosphere circulations have advected downwardly. \\
However, the best example of how significantly the zonal winds can affect the meridional circulations is SDSS1411B. Here, we see an almost completely different circulation profile than that found in the models of Kepler-13Ab, KELT-1b, or even HD209458b{. Instead, it is} much more similar to the { profile found }by \citet{2020ApJ...902...27T} for a rapidly rotating hot Jupiter atmosphere. The narrow equatorial jet, along with the dynamics that drive it, lead to a meridional circulation profile that is dominated by relatively weak narrow circulations cells, which do drive a downflow at the equator, while also driving significant (cooling) low latitude upflows{. This circulation structure may} help to explain why SDSS1411B struggles to maintain an advective adiabat { even when very weak radiative effects are included in }the deep atmosphere (see \autoref{sec:radiative_stability}).

\subsubsection{Influence of rotation on the flows and circulations}
{ Thus, we consider what is driving the above differences in flows and circulations. One possible explanation, other than the day-night temperature contrast, is the relative influence of rotational effects when compared to advective dynamics.}

To explore this balance, we  make use the of the Rossby number. This is a measure of the relative strength of advective forces to Coriolis accelerations:
\begin{equation}
R_{0} = \frac{U_{z}}{2\Omega\sin\left(\theta\right)L},
\end{equation}
where $U_{z}$ is the zonal velocity, $\Omega$ is the angular rotation rate, and $L$ is the characteristic length scale of the flows. We{ note that we used the zonal velocity since we are interested in how rotational effects affect the zonal flows, and we approximated the planetary radius as the characteristic length scale ($L=R_{BD}$) since we are interested in the global scale effect.} When the Rossby number is very small ($Ro\ll1$), rotational effects dominate over the advective dynamics; whereas when the Rossby number is large ($Ro\gg1$), rotational effects play a minor role in the dynamical balance. \\
{ Calculating the Rossby number for our three main deep-convergence brown dwarf models reveals that, as expected, the Rossby number near the equator (i.e. in the jet) tends to be significantly greater than 1: }at 0.01 bar, $R_{0} = 16.6$ for Kepler-13Ab, $R_{0} = 28.8$ for KELT-1b, and $R_{0} = 22.9$ for SDSS1411B. However, the Rossby number then drops rapidly as we move to higher latitudes, especially for SDSS1411B: at mid latitudes (i.e. around $45^\circ$), typical Rossby numbers are $R_{0} = 0.127$ for Kepler-13Ab, $R_{0} = 0.130$ for KELT-1b, and $R_{0} = 0.012$ for SDSS15411b{. The latter is similar to off-equator Rossby numbers found by} \citealt{2020ApJ...902...27T} for a rapidly rotating HD209458b-like model, and by \citealt{2020MNRAS.496.4674L} in models of WD0137-349B, another brown Dwarf that is in a close orbit around a white dwarf.\\
This suggests that off-equator flows are much more susceptible to rotational effects in SDSS1411B than in either { Kepler-13Ab or KELT-1b, which may help to} explain the very different zonal wind and meridional circulations profiles seen here.  We intend to explore the effects of changing rotation rate on the atmospheric dynamics, and the resulting deep steady-state, for highly irradiated exoplanets in much more detail as part of a future study.

\subsubsection{Exploring the vertical enthalpy and radiative fluxes}
{ To understand how the differences in flows and circulations affect the formation of a deep advective adiabat, we next explore the vertical advective heating rate and compare it to the deep radiative flux. To achieve this, we turn to the vertical enthalpy flux }
\begin{equation}
F_{H}= \rho c_{p} T U_{r}\,,
\end{equation}
where $U_{r}$ is the vertical velocity component.
For KELT-1b, we find a mean vertical enthalpy flux of $-1.11\times10^{9}\mathrm{\,erg\,s^{-1} cm^{-2}}$, which is significantly higher than the average deep radiative flux of $5.4\times10^{6}\mathrm{\,erg\,s^{-1} cm^{-2}}$. Similarly, for Kepler-13Ab, we find a mean vertical enthalpy flux of $-2.68\times10^{9}\mathrm{\,erg\,s^{-1} cm^{-2}}$, which is again significantly higher than the { average deep} radiative flux of $2.86\times10^{7}\mathrm{\,erg\,s^{-1} cm^{-2}}$. However, the same cannot be said for SDSS1411B, where we find a mean vertical enthalpy flux of $-4.92\times10^{6}\mathrm{\,erg\,s^{-1} cm^{-2}}$, which is almost equivalent to { average deep} radiative flux of $2.49\times10^{6}\mathrm{\,erg\,s^{-1} cm^{-2}}$. { We note that for all comparisons above, the deep radiative flux is taken from a 1D radiative-convective model that closely matches the 3D models' hot advective adiabat. \\}
This  explains both why the deep adiabat in our 100 bar convergence pressure model of SDSS1411B is so slow to heat and why it is so sensitive to deep radiative effects, namely: the highly rotationally influenced flows that develop in SDSS1411B lead to a significantly reduced vertical advective heating rate{. The result is a heating rate that is too slow to significantly heat the deep atmosphere and, hence, to cause radius inflation}. \\ 
The vertical enthalpy flux also helps to explain why the shallower convergence pressure models of Kepler-13Ab and KELT-1b have been shown to be more{ sensitive to deep radiative effects than models run at a deeper convergence pressures. For both Kepler-13Ab and KELT-1b,} not only does the slightly shallower jet drive slightly weaker equatorial downflows, but the shallower convergence pressure means that the average deep radiative flux, particularly near the convergence pressure, is higher. As such, it is harder for the model atmospheres to maintain an advective adiabat when deep radiative effects are included. For KELT-1b, our 10 bar convergence pressure model has a mean vertical enthalpy flux of $-1.07\times10^{8}\mathrm{\,erg\,s^{-1} cm^{-2}}$ and an average deep radiative flux of $1.86\times10^{7}\mathrm{\,erg\,s^{-1} cm^{-2}}$. Similarly, for Kepler-13Ab, our 1 bar convergence pressure model has a mean vertical enthalpy flux of $-2.21\times10^{9}\mathrm{\,erg\,s^{-1} cm^{-2}}$ and an average deep radiative flux of $\sim 1.1\times10^{8}\mathrm{\,erg\,s^{-1} cm^{-2}}$. \\
{ A similar effect has been found for test models of SDSS1411B with shallower horizontal convergence pressures, further reinforcing our analysis and conclusion that SDSS1411B should not be significantly inflated relative to 1D models. \\ }

\section{Discussion and conclusions} \label{sec:conclusion}
{ In this work, we present a series of models of Kepler-13Ab, KELT-1b, and SDSS1411B,  designed with the explicit purpose of exploring whether the observed radius inflation relation for brown dwarfs can be understood using the same mechanism that was shown to be applicable to hot Jupiters \citepalias{2019A&A...632A.114S,2017ApJ...841...30T}. 
That is the }vertical advection of potential temperature from a highly irradiated outer atmosphere to the deep atmosphere, leading to the formation of a hot and deep advective adiabat that is at a higher entropy than 1D `radiative-convective' models suggest (and, hence, is inflated { relative to the  1D models}).  \\

However, compared to the hot Jupiter explored by \citetalias{2019A&A...632A.114S} (HD209458b), the thermal structures of the outer atmospheres of the brown dwarfs under consideration here are not as well-documented. This presented the first major hurdle to this work as an accurate parametrisation of the outer atmosphere temperature-pressure (T-P) profile, including the day-night temperature difference at all pressures, is key to the Newtonian cooling approach (see \autoref{sec:method}) to radiative dynamics used here. An approach that remains necessary as the next generation of high performance, exascale, radiative GCMs, such as DYNAMICO-LMDZ\footnote{Which couples DYNAMICO with LMDZ: \url{https://lmdz.lmd.jussieu.fr}} or LFRic\footnote{Which is a project to develop a next generation replacement for the highly advanced MetOffice Unified Model: \url{https://www.metoffice.gov.uk/research/news/2019/gungho-and-lfric}}, are still in active development.

{ Although the observations are able to explore the day-night temperature contrast at very low pressures, they do not provide details about the full vertical structure of the T-P profile. In particular, observations do not help to constrain the horizontal convergence pressure at which horizontal homogenisation by advective flows and circulations has reduced the day-night temperature difference to zero. In \autoref{sec:free}, we demonstrate that freely varying the horizontal convergence pressure in models with no deep (i.e. $P>P_{converge}$) radiative forcing allows us to form an advective adiabat at almost any pressure (e.g. \autoref{fig:KELT_T_P_comp}). This includes} advective adiabats that were hot enough to explain either the observed radius or observed atmospheric features (\autoref{sec:obs}) of KELT-1b or Kepler-13Ab respectively (see \autoref{fig:obs_PT_comp}). 

{ Nonetheless, while the above models easily allow us to reproduce observations, they do not reveal any information about the true properties of the deep atmospheres of our brown dwarfs. After all, we can generate almost any deep advective adiabat in these no deep radiative forcing models just by adjusting the horizontal convergence pressure, and we cannot just run a single shallow convergence pressure model with deep radiative forcing because (as we discuss in \autoref{sec:flows}) the location of the day-night convergence affects the deep heating rate.} 

{ In order to try and resolve this issue, in \autoref{sec:1D_1000b} we next explored a series of models designed to remove any assumptions about where the day-side and night-side temperature profiles converge. More specifically, we ran a series of models with Newtonian cooling profiles based on the combination of an equilibrium 1D model paired with the observed day/night temperature difference in the outer atmosphere, and a deep atmosphere profile which used a sub-stellar-point 1D model to represent its day-side and a non-irradiated 1D model to represent its night-side.} For SDSS1411B{, both the 1D models, as well as our 3D models,} confirm that the deep atmosphere becomes both latitudinally and longitudinally converged at around 100 bar{. that is,} at essentially the same pressure that a convective adiabat is expected to develop in all of its fixed $T_{int}$ 1D models. { On the other hand, while the 1D models of both Kepler-13Ab and KELT-1b (\autoref{fig:1000b_multilong_lat}) do not converge at any pressure, the same cannot be said for the 3D models. These models} reveal that the atmosphere starts to converge longitudinally (and more slowly longitudinally) at 1 bar and 10 bar for Kepler-13Ab and KELT-1b, respectively, with the profile almost completely (longitudinally) converging at around $4.3$ bar and $40$ bar, respectively. We{ note, however, that although these models can help us to identify where the deep atmosphere convergences, the use of a somewhat unphysical deep Newtonian cooling profile with an imposed deep horizontal temperature gradients prevents us from using them to explore radius inflation in general.} 

{ As such, given that we already knew that advective adiabats could form within these convergence pressure ranges when we excluded radiative dynamics from the deep atmosphere (\autoref{sec:free}), we next explored (in \autoref{sec:radiative_stability}) what happens when this truncation is no longer in effect. To that end, we ran a series of models of all three brown dwarfs with horizontal convergence pressures based those found above, outer atmosphere forcing based on 1D equilibrium models and the day-night temperature difference, along with varying levels of deep radiative forcing towards an isotherm.}\\
{ Starting with Kepler-13Ab, we found that even as we vary the strength of the forcing, all test models with a shallower convergence pressure (1 bar) are highly sensitive to the presence{ of deep radiative effects}. However, this is not the case for models with a slightly deeper convergence pressure (4.3 bar). Here, while some cooling compared to the no-deep-forcing models does occur, the majority of our models maintain advective adiabats that are hotter than the convective adiabats found in our 1D models. Furthermore, it is important to note that this deep isothermal forcing is likely to over-exaggerate the impact of radiative cooling on the deep atmosphere -- test models with forcing towards the 1D models' convective adiabat revealed the same trend seen above, albeit to a much weaker extent, which is harder to quantify. As such, it is clear that potential temperature advection from the highly irradiated outer atmosphere to the deep atmosphere can provide an explanation of at }least some of the observed radius inflation of Kepler-13Ab. Furthermore it may, depending upon the exact nature of the deep radiative dynamics, be able to explain the observed atmospheric water features \citep{2017AJ....154..158B}. \\ 
A similar story holds true for { KELT-1b. Here,} models with a shallower convergence pressure (10 bar) fall into one of two regimes depending upon the exact form that the radiative timescale profile takes. Models with forcing profiles that include the region of near-constant $\tau_{rad}$ between $\sim10$ and $\sim100$ bar (\autoref{fig:PT_and_timescale}) generally exhibit significant deep cooling resulting in an advective adiabat that is notably cooler than the 1D models' convective adiabat{.
On the other hand,} model instead extrapolate the 1 bar timescale profile { down to 1000 bar} tend to exhibit much weaker cooling{. This results} in advective adiabats that remain hot enough to almost fully explain the { observed radius} inflation.{ Furthermore, moving onto models with a deeper convergence pressure (40 bar), we found that the importance of the form that the radiative timescale profile takes has been diminished. Instead, all the 40 bar deep forcing models explored here retain advective adiabats that are, at most,} only slightly cooled compared to their initial state. Thus, as is the case for Kepler-13Ab above, it is clear that potential temperature advection can explain at least some of the observed radius inflation of KELT-1b. { And in the right circumstances can explain all of the observed inflation.}\\
{ Finally we looked at the stability of an advective adiabat in a model of SDSS1411B with a 100 bar convergence pressure. Here we found that, for all but the weakest of deep radiative forcing, any initial adiabat rapidly cools} such that, after only 10 Earth years of simulation time, the deep, isothermally forced, T-P profile is already significantly cooler { than a 1D model's convective adiabat.} This suggests that the vertical advection of potential temperature does not lead to any significant radius inflation for SDSS1411B. This result with agrees with both the observations of SDSS1411B and brown dwarfs orbiting white dwarfs in general (\autoref{fig:MR_colour_2} and \autoref{sec:obs}). \\

To answer the question of why this difference in deep advective heating occurs, we next looked at the flows and circulations that drive this heating (\autoref{sec:flows}).
Here, we once again found that our brown dwarf models fall into one of two distinct regimes:\ the zonal wind and meridional circulation profiles of both Kepler-13Ab and KELT-1b behave very similarly to that of HD209458{; that is to say, we find a} super-rotating zonal jet that drives a strong equatorial downflow{ which, in turn, drives a strong downward enthalpy flux} ($-1.11\times10^{9}\mathrm{\,erg\,s^{-1} cm^{-2}}$ for KELT-1b and $-2.68\times10^{9}\mathrm{\,erg\,s^{-1} cm^{-2}}$ for Kepler-13Ab). \\
On the other hand, SDSS1411B reveals a very different jet structure { that consists of narrow alternating flows that weaken as they move to high latitudes.  This zonal wind profile drives a very different circulation profile than seen in either Kepler-13Ab or KELT-1b, which consists of very horizontally-narrow and vertically-extended circulation cells that drive a weak net downward enthalpy flux of a strength on the order of the deep radiative flux ($-4.92\times10^{6}\mathrm{\,erg\,s^{-1} cm^{-2}}$ versus $2.49\times10^{6}\mathrm{\,erg\,s^{-1} cm^{-2}}$), thus reducing the ability of advection to overcome radiative effects and heat the deep atmosphere of SDSS1411B.}

These differences in the flows and circulations are likely linked to the relative strength of mid to high latitude rotational effects in the two regimes{.  Analyses of the zonal }flows reveal that the mid-latitude Rossby number of SDSS1411B ($R_{0} = 0.012$) is an order of magnitude smaller than that of either KELT-1b or Kepler-13Ab ($R_{0} \sim 0.13$), suggesting that, for SDSS1411B, off-equator flows are significantly affected by the Coriolis effect, reducing the relative efficiency of advective heat transport. \\
A similar affect, albeit with a slightly different underlying cause, is also observed in the convective zones of stars: as the rotation rate is changed, the influence of rotation on the angular momentum flux balance changes{. As a result, the} meridional circulation profile shifts from a single cell in each hemisphere (slower, anti-solar, rotation) to multiple-cells aligned with the rotation axis (faster, solar-like, rotation -- here, cells become aligned with the rotation axis thanks to the effects of convection and the Taylor-Proudman constraint - \citealt{doi:10.1098/rspa.1917.0007,doi:10.1098/rspa.1916.0026}). This means that there is a similar transition in circulation dynamics at play as what is seen here; see, for example, \citet{2005LRSP....2....1M} or \citet{2011IAUS..271..261M} for a review of meridional circulation in solar convective zones. \\
The above findings suggest that the differences in flows and circulations that likely occur due to the differences in rotational influence between the slower rotating brown dwarfs with a main-sequence host star and the much more rapidly rotating brown dwarf orbiting a white dwarf can help to explain why the former show significant radius inflation and the latter do not. This simple, physical, explanation for the differences in brown dwarf inflation rates reinforces the idea that irradiation-induced and rotationally influenced advection of potential temperature appears to be a robust mechanism for resolving the radius-inflation problem in both hot Jupiters (\citetalias{2019A&A...632A.114S}) and brown dwarfs. \\

To further confirm that this is the case, we suggest that this work should be followed by two critical future studies. The first should explore how rotational effects influence the vertical advection of potential temperature and verify that this mechanism shuts down (or at least  significantly slows down) in highly rotationally influenced atmospheres. The second should, once the next generation of high performance (i.e. exascale) radiative GCMs are available, explore the steady state deep atmospheres of brown dwarfs models that include self-consistent radiative dynamics (and, hence, make fewer assumptions about the outer atmospheres' temperature structure, including the convergence pressure and the day-night temperature difference) and, thus, are able to obtain  a more reliable measurement of how much of the observed radius inflation can be adequately explained by advective heating. Additionally we suggest that WD1032b \citep{2020MNRAS.497.3571C} should be included in future studies of the radius inflation of brown dwarfs in order to investigate whether the slightly lower irradiation (cooler host star) and rotational influence (slower rotation) in the atmosphere can explain why this brown dwarfs orbiting a white dwarf might exhibit radius inflation when SDSS1411B does not. 


\begin{acknowledgements}
F. Sainsbury-Martinez and P. Tremblin would like to acknowledge and thank the ERC for funding this work under the Horizon 2020 program project ATMO (ID: 757858). 
The authors also wish to thank Idris, CNRS, and MDLS for access to the supercomputer Poincare, without which the long-timescale calculations featured in this work would not have been possible. Additionally this work was granted access to the HPC resources of IDRIS (Jean-Zay) and CEA-TGCC (Irene/Joliot-Curie) under the 2020/2021 allocation - A0080410870 made as part of the GENCI Dari A8 call.\\
Finally the authors with to thank the referee (and editor) for useful comments, questions, and suggestions. 
\end{acknowledgements}

\bibliographystyle{aa} 
\interlinepenalty=10000
\bibliography{papers}

\end{document}